\documentclass[pre,aps,reprint]{revtex4-1}

\usepackage{natbib}

\newenvironment{landscape}{}{}
\usepackage{setspace}

\usepackage{amssymb}
\usepackage{amsbsy}
\usepackage{amsmath}
\usepackage[varg]{txfonts}

\usepackage{amsmath}

\DeclareMathAlphabet{\mathsf}{OT1}{phv}{b}{n}

\newcommand{\Vector}[1]{\ensuremath{\mathbf{#1}}}
\newcommand{\Tensor}[1]{\ensuremath{\mathsf{#1}}}

\newcommand{\rmd}{{\mathrm{d}}}

\newcommand{\Int}[3]{\int\limits_{#1}^{#2}\!\rmd#3\;}

\newcommand{\Exp}[1]{{\rm e}^{#1}}
\newcommand{\OrderOf}[1]{\ensuremath{{\mathcal O}\left(#1\right)}}
\newcommand{\modulus}[1]{\ensuremath{\left\lvert#1\right\rvert}}

\newcommand{\sqfrac}[2]{\sqrt{\frac{#1}{#2}}}

\newcommand{\crossVorg}{\ensuremath{%
         \setbox0=\hbox{$V$}
        V \kern-\wd0{\raise.3ex\hbox{$\relbar$}}}}

\newcommand{\crossVxx}[2]{%
	{\setbox0=\hbox{$#1#2V$}
         \setbox1=\hbox{$#1#2$}
         \setbox2=\hbox{$#1V$}
         \dimen1=\wd0
	 \advance\dimen1-\wd1
         \raise.2\ht0\hbox{$#1#2$}\kern-.4\wd0}}

\newcommand{\ie}{i.e.}

\newcommand{\degree}{\ensuremath{^\circ}}
\newcommand{\degC}{\ensuremath{\degree\mathrm{C}}}

\usepackage{pifont}

\usepackage{graphicx}

\usepackage[colorinlistoftodos]{todonotes}
\usepackage{verbatim}
\usepackage{latexsym}
\usepackage{textcomp}
\usepackage{bm}
\usepackage{longtable}
\usepackage{subfigure}

\usepackage{epsfig}
\usepackage{dcolumn}
\usepackage{hyperref}
\usepackage[capitalize]{cleveref}

\newcommand{\qed}{\nobreak \ifvmode \relax \else
      \ifdim\lastskip<1.5em \hskip-\lastskip
      \hskip1.5em plus0em minus0.5em \fi \nobreak
      \vrule height0.75em width0.5em depth0.25em\fi}

\newcommand{\fnu}{\ensuremath{\Vector{F}_{\mathrm{\nu}}}}
\newcommand{\CS}{\ensuremath{C_{\text{S}}}}

\newcommand{\hsf}{\ensuremath{h^{*}_{\mathrm{f}}}}

\newcommand{\Rgth}{\ensuremath{R_{\mathrm{g}}^{\theta}}}

\newcommand{\Rhth}{\ensuremath{R_{\mathrm{H}}^{\theta}}}
\newcommand{\Rg}{\ensuremath{R_{\mathrm{g}}}}
\newcommand{\Rv}{\ensuremath{R_{\eta}}}

\newcommand{\Na}{\ensuremath{N_{\mathrm{A}}}}

\newcommand{\Rh}{\ensuremath{R_{\mathrm{H}}}}

\newcommand{\ag}{\ensuremath{\alpha_{\mathrm{g}}}}
\newcommand{\ah}{\ensuremath{\alpha_{\mathrm{H}}}}

\newcommand{\aeta}{\ensuremath{\alpha_{\mathrm{\eta}}}}

\newcommand{\Uer}{\ensuremath{U_{\mathrm{\eta R}}}}
\newcommand{\etas}{\ensuremath{\eta_{\mathrm{s}}}}

\newcommand{\etasp}{\ensuremath{\eta_{\mathrm{sp}}}}
\newcommand{\etapo}{\ensuremath{\eta_{\mathrm{p,0}}}}
\newcommand{\etao}{\ensuremath{\eta_{\mathrm{0}}}}

\newcommand{\ivisc}{\ensuremath{[\eta]}}
\newcommand{\etac}{\ensuremath{[\eta]_{\mathrm{c}}}}
\newcommand{\kh}{\ensuremath{k_{\mathrm{H}}}}
\newcommand{\kk}{\ensuremath{k_{\mathrm{K}}}}
\newcommand{\ksc}{\ensuremath{k_{\mathrm{SC}}}}

\newcommand{\Dth}{\ensuremath{D^{\theta}}}

\newcommand{\Ntoinf}{\ensuremath{N\to\infty}}
\newcommand{\hsH}{\ensuremath{h^{*}}}
\newcommand{\zs}{\ensuremath{z^{*}}}

\newcommand{\Uerinf}{\ensuremath{U_{\mathrm{\eta R}}^{\infty}}}

\DeclareMathAlphabet\mathbfcal{OMS}{cmsy}{b}{n}
\DeclareMathAlphabet{\mathbfsf}{\encodingdefault}{\sfdefault}{bx}{sl}

\DeclareMathOperator{\tr}{tr}

\newcommand{\bcal}[1]{\mathcal{#1}}

\newcommand{\bsf}[1]{\Tensor{#1}}

\newcommand{\bOmega}{\ensuremath{\boldsymbol{\varOmega}}}
\newcommand{\bdelta}{\ensuremath{\boldsymbol{\delta}}}

\newcommand{\br}{\ensuremath{\bm{\Vector{r}}}}
\newcommand{\brhat}{\ensuremath{\bm{\Vector{\hat r}}}}

\newcommand{\erf}{\text{erf}}

\begin{document}

\author{Sharadwata Pan}
\affiliation{IITB-Monash Research
  Academy, Indian Institute of Technology Bombay, Powai, Mumbai
  400076, India; Department of Chemical Engineering, Indian
  Institute of Technology Bombay, Powai, Mumbai 400076, India and Department of Chemical Engineering, Monash University,
  Melbourne, VIC 3800, Australia}

\author{Deepak Ahirwal}
\affiliation{Department of Chemical Engineering, Indian
  Institute of Technology Bombay, Powai, Mumbai 400076, India}

\author{Duc At Nguyen}
\affiliation{Department of Chemical Engineering, Monash University,
  Melbourne, VIC 3800, Australia}

\author{P. Sunthar}
\affiliation{Department of Chemical Engineering, Indian Institute of Technology Bombay, Powai, Mumbai 400076, India and IITB-Monash Research Academy, Indian Institute of Technology Bombay, Powai, Mumbai 400076, India}

\author{T. Sridhar}
\affiliation{Department of Chemical Engineering, Monash University,
  Melbourne, VIC 3800, Australia and IITB-Monash Research Academy, Indian Institute of Technology Bombay, Powai, Mumbai
  400076, India}


\author{J. Ravi Prakash}
\thanks{Communicating Author}
\email{ravi.jagadeeshan@monash.edu}
\affiliation{Department of Chemical Engineering, Monash University,
  Melbourne, VIC 3800, Australia and IITB-Monash Research Academy, Indian Institute of Technology Bombay, Powai, Mumbai
  400076, India}

\title[Viscosity radius of DNA solutions]{The viscosity radius of polymers in dilute solutions: Universal behaviour from DNA rheology and Brownian dynamics simulations}



\begin{abstract}

The swelling of the viscosity radius, $\aeta$, and the universal viscosity ratio, $\Uer$, have been determined  experimentally for linear DNA molecules in dilute solutions with excess salt, and numerically by Brownian dynamics simulations, as a function of the solvent quality. In the latter instance, asymptotic parameter free predictions have been obtained by extrapolating simulation data for finite chains to the long chain limit. Experiments and simulations show a universal crossover for \aeta\ and \Uer\ from $\theta$ to good solvents in line with earlier observations on synthetic polymer-solvent systems. The significant difference between the swelling of the dynamic viscosity radius from the observed swelling of the static radius of gyration, is shown to arise from the presence of  hydrodynamic interactions in the non-draining limit. Simulated values of \aeta\  and \Uer\  are in good agreement with experimental measurements in synthetic polymer solutions reported previously, and with the measurements in linear DNA solutions reported here.
\end{abstract}

\maketitle

\section{\label{sec:intro} Introduction}

Large scale static and dynamic properties of dilute polymer solutions scale as power laws with molecular weight $M$ in the limits of $\theta$ and very good solvents~\cite{dgen79,RubCol03}. In the intermediate regime between these two limits, their behaviour can be described in terms of \textit{crossover} functions of a single scaling variable, the so-called solvent quality parameter, $\tilde z = \frac{3}{4} K(\lambda L) \, z$, where $K$ is a function of the chain stiffness parameter ($\lambda^{-1}$) and contour length ($L$), and the parameter $z = k (1 - T_{\theta}/T)\sqrt{M}$, combines the dependence on temperature $T$ and molecular weight~\cite{Yama01,yamakawa1997,Schafer99}. The constant $k$ is chemistry dependent, and $T_{\theta}$ is the $\theta$-temperature. In the random coil limit $\lambda L \to \infty$, where polymer chains are completely flexible, $\tilde z = z$. Examples of such crossover functions include the swelling functions, $\ag\ = \Rg / \Rgth$ (which is a ratio of the radius of gyration at any temperature $T$ to the radius of gyration at the $\theta$-temperature), $\ah\ = \Rh / \Rhth$ (where \Rh\ is the hydrodynamic radius), and $\alpha_{\eta} = {R_{\eta}}/{R_{\eta}^{\theta}}=\left({[\eta]}/{[\eta]_{\theta}}\right)^{{1}/{3}}$, where $R_{\eta}$ is the viscosity radius, defined by the expression,
\begin{equation}
\label{eq:RV}
R_{\eta} \equiv \left(\frac{3 [\eta] M }{10 \pi N_{A}}\right)^{{1}/{3}}
\end{equation}
with \Na\ being the Avogadro's constant, and \ivisc\ the zero shear rate intrinsic viscosity. Several experimental studies~\cite{MiyFuj81,Arai1995,Tominaga20021381,Hayward19993502} have shown that swelling data for many different polymer-solvent systems, can be collapsed onto master plots, independent of chemical details, when represented in terms of the parameter $\tilde z$. Notably, however, the universal curve for \ag\ (which is a ratio of static properties), is significantly different from the universal curves for \ah\ and \aeta, which are ratios of dynamic properties~\cite{MiyFuj81,Arai1995,Tominaga20021381,Hayward19993502}. There have been many attempts to understand the origin of this difference in crossover behaviour, and to predict analytically and numerically, the observed universal curves~\cite{Weill197999,Benmouna19781187,Douglas19842344,Douglas19842354,Dunweg2002914,Yamakawa1995,YoshiYama96,yamakawa1997,Freed1988,SunRav06-epl} (see  Ref.~\citenum{Jamieson2010} for a recent review). In this paper, we re-examine this problem in the context of Brownian dynamics (BD) simulations, which are a means of obtaining an exact (albeit numerical) solution to the underlying model for the polymer solution. We also report on experimental measurements of the viscosity radius of DNA in the presence of excess salt (at two different molecular weights), and examine the universality of the crossover of properties derived from the viscosity radius by comparison with previous measurements for synthetic polymer solutions.

Dilute polymer solution models typically represent polymers as chains of beads connected together by rods or springs, immersed in a Newtonian solvent. The beads act as centres of frictional resistance to chain motion through the solvent, and the motion of all the beads are coupled together through hydrodynamic interactions which represent the solvent mediated propagation of momentum between the beads. Bead overlap is prevented either by excluded volume interactions between the beads, acting  pair-wise through a repulsive potential, or through restriction of chain configurations to those that are self-avoiding. Within such a framework, analytical theories such as the renormalisation group theory~\cite{Schafer99} and two-parameter theories~\cite{Yama01} have successfully predicted static properties of dilute solutions of flexible polymers. For instance, both renormalisation group and two-parameter theories provide explicit expressions for \ag\ as a function of $z$. A well known example of the latter is the Domb-Barrett equation~\cite{dombbarrett76,Jamieson2010}.

Both the hydrodynamic and viscosity radii are dynamic properties, and consequently, hydrodynamic interactions play a crucial role in determining the swelling functions \ah\ and \aeta. Barrett~\cite{Barrett1984} used two-parameter theory with \emph{pre-averaged} hydrodynamic interactions to develop   explicit expressions for \ah\ and \aeta\ as functions of $z$. The Barrett equation for \aeta\ has proven to be an extremely accurate means of predicting the swelling of $R_{\eta}$  for a number of different polymer-solvent systems~\cite{Tominaga20021381,yamakawa1997,Jamieson2010}. On the other hand, the Barrett equation for \ah\ considerably over-predicts the extent of swelling of the hydrodynamic radius when compared to experimental measurements in the crossover regime~\cite{Arai1995,Tominaga20021381,yamakawa1997}. Zimm~\cite{Zimm1980} first recognised that the neglect of fluctuating hydrodynamic interactions in models with pre-averaged hydrodynamic interactions could be a significant factor responsible for the poor prediction of universal properties. Yamakawa and coworkers~\cite{Yamakawa1995,YoshiYama96,yamakawa1997} subsequently developed an approximate analytical model to account for the presence of fluctuating hydrodynamic interactions within the framework of quasi-two-parameter theory, which is a modification of two-parameter theory that accounts for chain stiffness by introducing the parameter $\tilde z$ in place of $z$ as the universal scaling variable. They suggest that
$ \ah = \ah^{(0)} \, h_{H}$, and $\aeta = \aeta^{(0)} \, h_{\eta}$, where $ \ah^{(0)}$ and  $\aeta^{(0)}$ are the swelling functions predicted in the absence of fluctuations, and $h_{H}$ and $h_{\eta}$ are corrections that account for their presence. \citet{Yamakawa1995} have proposed an expression for $h_{H}$ as a function of $\tilde z$, while currently there is no analytical expression for $h_{\eta}$. The inclusion of fluctuations in hydrodynamic interactions in this manner leads to a reduction in the values of \ah\ predicted by the Barrett equation, however, they are still too high relative to experimental values in the entire crossover regime~\cite{Arai1995,Tominaga20021381,yamakawa1997}.

An alternative explanation~\cite{Freed1988,Jamieson2010} that has been offered for the difference in the universal crossover functions for \ah\ and \aeta\ from \ag, is that hydrodynamic interactions are not fully developed in the crossover regime, i.e., rather than being in the \emph{non-draining} limit where polymer coils behave as rigid spheres, there is a \emph{partial-draining} of the solvent through polymer coils, which are {swollen} because of excluded volume interactions. This approach, however, also does not result in an improved prediction of the universal crossover function for \ah~\cite{SunRav06-epl}.

More recently,~\citet{SunRav06-epl} have shown for flexible polymer chains, by carrying out exact BD simulations of bead-spring chains, that the difference between \ag\ and \ah\ is in fact due to the presence of fluctuating hydrodynamic interactions in the non-draining limit. By accounting for fluctuating hydrodynamic and excluded volume interactions in the asymptotic long chain limit, Prakash and coworkers have been able to obtain quantitatively accurate, parameter free predictions of \ag\ and \ah\ as functions of $z$~\cite{Kumar20037842,SunRav06-epl}.

The agreement of the Barrett equation~\cite{Barrett1984} for \aeta\ with experimental observations has been taken to imply that, in contrast to \ah, fluctuations in hydrodynamic interactions are not important in determining the swelling of the viscosity radius~\cite{YoshiYama96,yamakawa1997}. However, this cannot be conclusively established without an exact estimate of the magnitude of fluctuations in the entire crossover regime. For instance, the agreement could arise fortuitously from a cancellation of errors due to the assumption of pre-averaged hydrodynamic interactions and the occurrence of partial-draining. The use of BD simulations provides an opportunity to account exactly for the presence of fluctuating hydrodynamic interactions, and consequently, to examine its role in determining the observed difference in the crossover of  \aeta\ and \ag, as has been done previously  in the case of \ah\ by~\citet{SunRav06-epl}.

Properties of dilute polymer solutions are often measured in order to obtain structural information about the dissolved macromolecules. By comparing experimental data with predictions of solution models  with different macromolecular structures, such as flexible, wormlike, ellipsoidal, cylindrical, etc., information on the shape, size and flexibility of macromolecules can be obtained. Rather than using the values of properties themselves, it has been found more convenient to construct dimensionless ratios of properties, since such ratios tend to depend only on the shape of the macromolecule, and not on its absolute size. A well known example of such a ratio, based on the intrinsic viscosity and radius of gyration, is the Flory-Fox constant~\cite{RubCol03}, $\Phi =  [\eta] M /6^{\frac{3}{2}} \Rg^{3}$. An alternative approach proposed by Garc\'{\i}a de la Torre and coworkers, is to use equivalent radii instead of properties themselves to construct dimensionless ratios~\cite{Torre07,Torre11}. An equivalent radius is defined as the radius of a sphere, a dilute suspension of which would have the same value of the property as the solution itself. For instance, $R_{\eta}$ defined by~\cref{eq:RV}, and $\text{GI} = \sqrt{5/3} \, (\Rg/R_{\eta})$,  are examples of an equivalent radius and a non-dimensional ratio of equivalent radii, respectively. Garc\'{\i}a de la Torre and coworkers have shown that the use of such ratios is a more efficient and less error prone way of extracting structural information~\cite{Torre07,Torre11}.

We use the viscosity ratio, $U_{\eta R}$, which is usually defined in the context of BD simulations~\cite{ottinger,Kroger20004767}, as a universal function that characterises polymer solutions. It is trivially related to both $\Phi$ and $\text{GI}$,
\begin{equation}
\label{eq:uetar}
U_{\eta R} \equiv \frac{5}{2} \left( \frac{\Rv}{\Rg} \right)^{3} = \frac{6^{\frac{3}{2}}}{(4\pi/3)}\, \frac{\Phi}{\Na} =  \frac{5}{2} \left( \frac{5}{3} \right)^{\frac{3}{2}} \left( \text{GI}\right)^{-3}
\end{equation}
\citet{Kroger20004767} have tabulated experimentally measured values of $U_{\eta R}$, and the predictions of various approximate theories and simulations (under both $\theta$ solvent and good solvent conditions). For $\theta$ solvents, experimental measurements~\cite{Miyaki1980} indicate that $U_{\eta R}^{\theta} = 1.49 \pm 0.06$, which corresponds to the well known value of the Flory-Fox constant for flexible polymers in $\theta$ solvents, $\Phi_{0} = 2.56 \times 10^{23}$. Garc{\'\i}a de la Torre and coworkers~\cite{Torre84,Torre86,Torre91,Torre11} have used the Monte Carlo rigid body method, accompanied by extrapolation of finite chain data to the long chain limit, to  predict $\Phi_{0} = 2.53 \times 10^{23}$  in $\theta$ solvents (which equates to~\cite{Kroger20004767}  $U_{\eta R}^{\theta} \approx1.47 \pm 0.15$), while in the limit of very good solvents ($z \to \infty$) they predict, $\Phi = 1.9 \times 10^{23}$  (i.e., $U_{\eta R}^{\infty}\approx1.11 \pm 0.10$). By carrying out non-equilibrium BD simulations at finite shear rates, and extrapolating the finite shear rate data to the limit of zero shear rate, \citet{Kroger20004767} predict $U_{\eta R}^{\theta} \approx1.55 \pm 0.04$. \citet{Jamieson2010} observe that even though a number of experimental measurements of the Flory-Fox constant under good solvent conditions have been reported in the literature, the behaviour of $\Phi$ with varying solvent conditions and molecular weight appears not to be understood with any great certainty.

An analytical expression for the crossover behaviour of the ratio $U_{\eta R}/U_{\eta R}^{\theta}$ (which is also equal to the ratio of the Flory-Fox constants in good and $\theta$-solvents) can be determined by substituting the Domb-Barrett equation~\cite{dombbarrett76} for \ag, and the  Barrett equation~\cite{Barrett1984} for \aeta\  in the right-hand-side of the expression below (which follows from the definitions of the various quantities involved),
\begin{equation}
\label{eq:uetaratio1}
\frac{U_{\eta R}}{{U_{\eta R}^{\theta}}} =
\left(  \frac{\alpha_{\eta}}{\ag} \right)^{3}
\end{equation}
Not surprisingly, given the accuracy of the Domb-Barrett and Barrett equations, experimental data on the crossover of this ratio is well captured by quasi-two-parameter theory~\cite{Tominaga20021381,Jamieson2010}. However, as in the case of the expansion factor \aeta,  so far  no exact Brownian dynamics simulations have been carried out to determine the crossover behaviour of $U_{\eta R}$ (a knowledge of which would also provide the ratio $U_{\eta R}/U_{\eta R}^{\theta}$).

Reported observations of \aeta\ and $\Phi$ have largely been on synthetic polymer-solvent systems~\cite{Tominaga20021381,MiyFuj81,Hayward19993502}. Recently, \citet{Pan2014339} have shown that the crossover swelling of the hydrodynamic radius of linear DNA molecules in dilute solutions with excess salt can be collapsed onto earlier observations of the swelling of the hydrodynamic radius of synthetic polymers. This result was established by: (i) showing with the help of static light scattering that the $\theta$-temperature of a commonly used excess salt solution of linear DNA molecules is $T_{\theta} \approx 15 \degC$,  and (ii) by estimating the hydrodynamic radius and the solvent quality at any temperature and molecular weight by dynamic light scattering measurements. These developments make it now possible to examine the crossover behaviour of any static or dynamic property of linear DNA solutions in the presence of excess salt.

The aim of this paper is two-fold: (i) To carry out systematic measurements of the intrinsic viscosity of two different molecular weight samples of linear double-stranded DNA at a range of temperatures in the presence of excess salt, and examine the crossover scaling of the swelling of the viscosity radius \aeta, and the viscosity ratio, $\Uer$. Comparison with earlier observations of the behaviour of synthetic polymers enables not only the establishment of the universal scaling of DNA solutions, but also serves as an independent verification of the earlier estimate of the $\theta$-temperature and solvent quality by \citet{Pan2014339} (ii) To carry out detailed BD simulations of bead-spring chains to estimate \aeta\ and $\Uer$ as functions of $z$ for flexible polymers. This has previously been difficult because of the large error associated with simulations of viscosity at low shear rates. By using a Green-Kubo formulation, and a variance reduction scheme, coupled with systematic extrapolation of finite chain data to the long chain limit to circumvent the problem of poor statistics, we show for the first time that by including fluctuating excluded volume and hydrodynamic interactions, quantitatively accurate prediction of the crossover scaling of \aeta\ and  $\Uer$ can be obtained, free from the choice of arbitrary model parameters. Further, the difference between the crossover scaling of \ag\ and \aeta\ is shown to arise undoubtedly from the influence of hydrodynamic interactions in the non-draining limit, and the relative unimportance of fluctuations in hydrodynamic interactions is  confirmed.

The plan of the paper is as follows. In~\cref{sec:expt} on ``Materials and Methods'', we describe the experimental protocol for preparing the DNA samples and for carrying out the viscosity measurements. We also discuss the governing equations for the BD simulations, the variance reduction scheme adopted here, and the calculation of the viscosity using a Green-Kubo expression. In~\cref{subsec:ivisc}, we describe the measurement of the intrinsic viscosity of the DNA solutions, and tabulate values of intrinsic viscosity and the Huggins coefficient across a range of temperatures. In the remaining subsections of~\cref{sec:resultexpt}, we discuss the prediction of \aeta\ and $\Uer$ by BD simulations, and compare simulation predictions with prior and current experimental measurements. Finally, in~\cref{sec:conc}, we summarise the principal conclusions of the present work.

\section{Materials and Methods}
\label{sec:expt}
\subsection{DNA samples and shear rheometry}

Viscosities have been measured for two different double stranded DNA molecular weight samples: (i) T4 bacteriophage linear genomic DNA [size 165.6 kilobasepairs (kbp)] and (ii) 25 kbp DNA. While the former were obtained from Nippon Gene, Japan (\#314-03973), the latter were extracted, linearized, and purified from \emph{Escherichia coli} (\emph{E. coli}) stab cultures procured from Smith's laboratory at UCSD. Smith's group have genetically engineered special double-stranded DNA fragments in the range of 3-300 kbp and incorporated them inside commonly used \emph{E. coli} bacterial strains. These strains can be cultured to produce sufficient replicas of its DNA, which can be cut precisely at desired locations to extract the special fragments~\cite{Laib20064115}. The protocol for preparing the 25 kbp samples obtained in this manner has been described in detail in \citet{Pan2014339} Typical properties of the DNA molecules used in this work, such as the molecular weight, contour length, number of Kuhn steps, etc., are tabulated in Table~S-1 (Supporting Information). Additionally, details regarding the solvent, and estimation of DNA concentration, etc., are presented in the Supporting Information.

A Contraves Low Shear 30 rheometer has been used to obtain all the shear viscosity measurements reported in the present work  because of two main advantages: it has a zero shear rate viscosity sensitivity even at a shear rate of 0.017 $s^{-1}$ and thus can measure very low viscosities; and has a very small sample requirement (minimum 800 $\mu$l)~\cite{Heo20051117}. Both of these are ideal for measuring viscosities of biological samples such as DNA solutions. The steady state shear viscosities $\eta$ were measured at low concentrations ($c < c^{*}$) and across a temperature range of 15--35\degC. The overlap concentrations ($c^{*}$), at different temperatures, were estimated from the known  values of the solvent quality parameter $z$, as described in \citet{Pan2014339} The zero shear rate viscosity was determined from measurements of viscosity at different finite shear rates, and extrapolation to zero shear rate. Details are given in the Supporting Information. Values obtained this way for the two molecular weights, across the range of concentrations and temperatures, are displayed in Table~S-2 (Supporting Information).

\subsection{\label{sec:BD} Brownian dynamics simulations}

The dilute polymer solution is modelled as an ensemble of non-interacting bead-spring chains, immersed in a Newtonian solvent. Each chain has $N$ beads of radius $a$, connected together by Hookean springs with spring constant $H$. The beads act as centres of frictional resistance, with a Stokes friction coefficient, $\zeta = 6 \pi \etas \, a$ (where $\etas$ is the solvent viscosity), and bead overlap is prevented through a pair-wise repulsive narrow Gaussian excluded volume potential (which is a regularisation of a delta function potential). Hydrodynamic interactions between the beads are modelled with the Rotne-Prager-Yamakawa (RPY) regularisation of the Oseen function. Within this framework, the time evolution of the positions of the $N$ beads, ${\Vector{r}}_{1}(t), {\Vector{r}}_{2}(t), \ldots, {\Vector{r}}_{N}(t)$, are governed by stochastic differential equations, which can be integrated numerically (exactly) with the help of Brownian dynamics simulations. Details of the stochastic differential equations, the precise forms of the excluded volume potential and hydrodynamic interaction tensor, and key aspects of the integration algorithm, are given in the Supporting Information. It is sufficient to note here that by adopting the length scale $l_H = \sqrt{k_B T/H}$ and time scale ${\lambda}_H = \zeta/4H$ for the purpose of non-dimensionalization (where $k_B$ is Boltzmann's constant), it can be shown that there are three parameters that control the dynamics of finite bead-spring chains at equilibrium, namely, the number of beads $N$, the strength of excluded volume interactions $z^{*}$, and the hydrodynamic interaction parameter, $\hsH = a \sqrt{H/(\pi k_B T)}$.

Analytical theories have shown that the true strength of hydrodynamic interactions is determined by the {draining} parameter,~\cite{Zimm56,ottrab89}  $h=h^{*} \sqrt{N}$, while, for flexible polymers, the strength of excluded volume interactions is determined by the excluded volume parameter,~\cite{Schafer99,prakash:macromol-01} $z = z^{*} \sqrt{N}$. Note that the experimentally measured solvent quality parameter defined previously for flexible chains can be mapped onto theoretical values of $z$ by a suitable choice of the constant $k$~\cite{Kumar20037842}.

Universal predictions, independent of details of the coarse-grained model used to represent a polymer, are obtained in the limit of long chains, since the self-similar character of real polymer molecules is captured in this limit. It is common to obtain predictions in the long chain limit by accumulating data for finite chain lengths and extrapolating to the limit $N \to \infty$. This procedure has been used successfully to calculate universal properties of dilute polymer solutions predicted by a variety of approaches to treating hydrodynamic and excluded volume interactions, including Monte Carlo simulations~\cite{Zimm1980,Torre84,Torre86,Torre91}, approximate closure approximations~\cite{ott87d,Ottinger:1989up,prakash:jnnfm-97,prakash:jor-02}, and exact Brownian dynamics simulations~\cite{Kroger20004767,Kumar20037842,kumar:jcp-04,SunRav05,SunRav06-epl,Bosko:2011wk}.

The non-draining limit corresponds to $h \to \infty$. As a result, simulations carried out at constant values of $h^{*}$ naturally lead to predictions in the non-draining limit as $N \to \infty$. \citet{SunRav06-epl} have shown that universal predictions in the non-draining limit, and at any fixed value of the solvent quality parameter $z= \zs \, \sqrt{N}$, can be obtained by simultaneously keeping $\hsH$ and $z$ constant, while  taking the limit $\Ntoinf$. Since the parameter $\zs \to 0$ in this limit, the long chain limit of the model corresponds to the Edwards continuous chain model with a delta function excluded volume repulsive potential~\cite{DoiEd86}. As mentioned in~\cref{sec:intro}, by accounting for fluctuating hydrodynamic and excluded volume interactions in this manner, \citet{SunRav06-epl} have obtained a quantitatively accurate parameter free prediction of \ah\ as a function of $z$. Here, we show that this approach can also be used to successfully predict universal properties related to the zero shear rate viscosity of dilute polymer solutions.

\subsection{\label{sec:uni} Universal properties derived from the viscosity radius}

We focus our attention on two properties that are defined in terms of
the viscosity radius (\cref{eq:RV}) which have been shown to be universal in the sense that they are independent of the chemistry of the particular polymer-solvent system for sufficiently long polymers. The first of these is the universal viscosity ratio, $U_{\eta R}$ (defined in~\cref{eq:uetar}), and the second is the swelling ratio $\alpha_{\eta}$. We discuss the evaluation of these properties by Brownian dynamics simulations in turn below.

In terms of dimensionless variables, $U_{\eta R}$ can be shown to be given by
\begin{equation}
\label{eq:uetarsim} U_{\eta R} = \frac{9}{8} \sqrt \pi \hsH
\frac{\etapo^{*}}{{\Rg^{*}}^{3}}
\end{equation} where, $\Rg^{*}$ is the dimensionless radius of gyration, and $\etapo^{*} = \etapo / (n_\text{p}\lambda_{H} k_B T)$ is the dimensionless zero-shear rate viscosity. Here, $n_\text{p}$ is the number of chains per unit volume, and $\eta_{\mathrm{p, 0}} = \etao-\etas$, is the polymer contribution to the zero shear rate solution viscosity.  \citet{Kroger20004767} have estimated $\etapo^{*}$ by
carrying out  non-equilibrium BD
simulations at finite shear rates, and extrapolating the data to the limit
of zero shear rate. Here, we use an alternative method based on a
Green-Kubo relation~\cite{Fixman1981} which gives the viscosity as an
integral of the equilibrium-averaged stress-stress auto-correlation
function
\begin{equation}
\label{eq:etap0gk} \etapo^{*} = \Int{0}{\infty}{t} \langle \CS ({\br}_{1}, {\br}_{2},\ldots, {\br}_{N}, t) \rangle_{\mathrm{eq}}
\end{equation}
where,
\begin{equation}
\label{eq:Sxyauto} \CS({\br}_{1}, {\br}_{2},\ldots, {\br}_{N},t) =  S_{xy}(t) S_{xy}(0)
\end{equation}
The quantity $S_{xy}$ is the $xy$-component of the stress tensor given by Kramers expression $S_{xy} =
\sum_{\mu} F_{\mu x} (r_{\mu y} - {r}_{cy})$,
where $r_{\mu y}$ is the $y$-component of $\br_{\mu}$, ${r_{cy}}$ is the $y$-component of the position vector of the center-of-mass of the bead-spring chain, $\br_{c} = (1/N) \sum_{\mu}  \br_{\mu}$, and  $F_{\mu x}$ is the $x$-component of  ${\bm{\Vector{F}}}_{\mu}$, the sum of all the non-hydrodynamic forces on bead $\mu$ due to all the other beads.  The use of the Green-Kubo method mitigates the problem of the large
error bars associated with estimating polymer solution properties at low shear rates. We find that the noise in measured properties can be significantly reduced by evaluating the integral in \cref{eq:etap0gk} with the help of equilibrium simulations of a large ensemble of trajectories.  Additionally, for some simulations, we have employed a variance reduction technique, as explained in~\cref{sec:var} below.

Rather than evaluating the swelling of the viscosity radius directly from its definition $\alpha_{\eta} = {R_{\eta}}/{R_{\eta}^{\theta}}$, we found it advantageous to use the following expression (obtained by rearranging \cref{eq:uetaratio1}), which gives $\alpha_{\eta}$ in terms of $U_{\eta R}$ and \ag, since the $\Ntoinf$ extrapolations of $U_{\eta R}^{\theta}$ and $U_{\eta R}$ (at various values of $z$) are more accurate than the extrapolations for $\alpha_{\eta}$,
\begin{equation}
\label{eq:aeta1} \alpha_{\eta} =
\left( \frac{U_{\eta R}}{{U_{\eta R}^{\theta}}}\right)^{{1}/{3}} \ag
\end{equation}
The swelling of the radius of gyration \ag\ for different values of $z$ is calculated from the expression,
\begin{equation}
\label{eq:ag} \ag = (1 + az + bz^{2} + cz^{3})^{m/2}
\end{equation}
with values of the fitting parameters, $a$, $b$, $c$, and $m$ as given in~\cref{tab:fitpars}. This specific form for the fitting function is often used in renormalization group theory predictions and in lattice simulations to represent the crossover behaviour of swelling ratios for flexible chains~\cite{Schafer99}.  \cref{eq:ag} has been shown by \citet{Kumar20037842} to be an excellent fit to the asymptotic predictions of \ag\ by BD simulations in the absence of hydrodynamic interactions. This corresponds to the pure excluded volume problem, which is adequate for determining \ag, since it is a static property unaffected by hydrodynamic interactions.

\subsection{\label{sec:var} Variance reduced simulations}

The statistical error in the estimation of the equilibrium-averaged stress-stress auto-correlation function $\langle \CS (t) \rangle_{\mathrm{eq}}$ can be significantly reduced if the fluctuations in $\CS({\br}_{1}, {\br}_{2},\ldots, {\br}_{N},t)$ can be made to be small. Amongst the many approaches available for reducing the magnitude of fluctuations in stochastic simulations~\cite{ottinger}, we have adopted a variance reduction technique based on the use of control variates~\cite{melott96}, as described below.

\begin{figure}[!tbp]
\centering
\resizebox{0.9\linewidth}{!} {\includegraphics{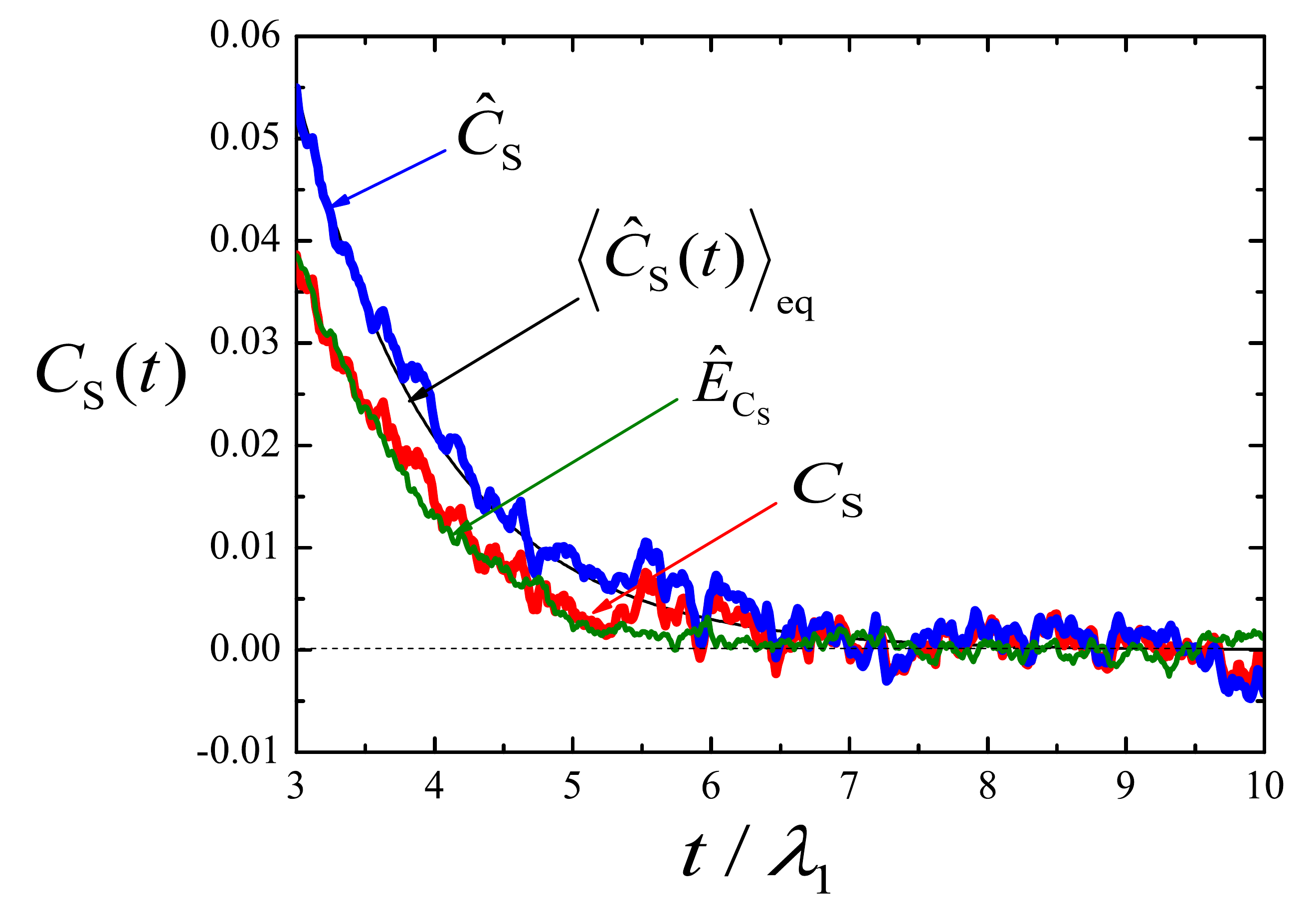}}
\caption{Reduction in the variance of the stress auto-correlation function. The two auto-correlation functions, \CS\ (red curve) calculated with fluctuating hydrodynamic interactions, and ${\hat \CS}$ (blue curve) calculated with pre-averaged hydrodynamic interactions, can be seen visually to be positively correlated. The control variate $\hat E_{\text{C}_\text{S}}$ (green curve), clearly has significantly lower fluctuations. The analytical function $\langle {\hat \CS (t)} \rangle_{\mathrm{eq}}$ (black curve) is given by Fixman's expression~\cite{Fixman1981} (see Supporting Information). The range of the axes have been chosen to magnify the noise at small values of \CS. In this simulation $\lambda_{1} = 38.2$, is the longest relaxation time, estimated from Thurston's correlation~\cite{Thurston1974569} for $N=18$, and $\hsH=0.25$. The averages have been obtained over roughly 57000 independent trajectories.}
\label{fig:vred}
\end{figure}

In general, the fluctuations, $f_{\text{C}_\text{S}} = \CS({\br}_{1}, {\br}_{2},\ldots, {\br}_{N},t) - \langle \CS (t) \rangle_{\mathrm{eq}}$, cannot be estimated a priori. However, if the fluctuations, $ {\hat f}_{\text{C}_\text{S}} = {\hat \CS} ({\brhat}_{1}, {\brhat}_{2},\ldots, {\brhat}_{N},t) - \langle {\hat \CS (t)} \rangle_{\mathrm{eq}}$, can be determined for a stochastic process $\brhat_{\nu}$ for which the equilibrium-averaged stress-stress auto-correlation $\langle {\hat \CS (t)} \rangle_{\mathrm{eq}}$ is known analytically, and $ {\hat f}_{\text{C}_\text{S}} \approx f_{\text{C}_\text{S}}$, then, the control variate
\begin{equation}
\label{eq:E}  \hat E_{\text{C}_\text{S}} =  \CS({\br}_{1}, {\br}_{2},\ldots, {\br}_{N},t) - {\hat f}_{\text{C}_\text{S}}
\end{equation}
can be used to estimate the stress-stress auto-correlation function with reduced statistical error, since $\langle \hat E_{\text{C}_\text{S}} \rangle_{\mathrm{eq}}  =  \langle \CS (t) \rangle_{\mathrm{eq}}$. The extent of the reduction in statistical error depends on the extent to which $C_\text{S}$ and $\hat C_\text{S} $ are correlated, as can be seen from the expression for the variance of $\hat E_{\text{C}_\text{S}}$,
\begin{align}
\label{eq:varE}
\left\langle \left[ \hat E_{\text{C}_\text{S}} -  \langle E_{\text{C}_\text{S}} \rangle_{\mathrm{eq}} \right]^{2} \right\rangle_{\mathrm{eq}} = \left\langle \left[ C_\text{S} -  \langle C_\text{S} \rangle_{\mathrm{eq}} \right]^{2} \right\rangle_{\mathrm{eq}} + \left\langle \left[ \hat C_\text{S} -  \langle \hat C_\text{S} \rangle_{\mathrm{eq}} \right]^{2} \right\rangle_{\mathrm{eq}}\nonumber\\
- 2 \left[  \langle  C_\text{S} \, \hat C_\text{S} \rangle_{\mathrm{eq}} - \langle C_\text{S}  \rangle_{\mathrm{eq}}  \langle \hat C_\text{S} \rangle_{\mathrm{eq}} \right]
\end{align}
We use the stochastic process $\brhat_{\nu}$, governed by the stochastic differential equation,
\begin{equation}
\rmd \brhat_{\mu} = \frac{1}{4} \, \sum_{\nu} H_{\mu\nu} \,  \fnu \,
\rmd t + \frac{1}{\sqrt{2}} \, \sum_{\nu} S_{\mu\nu} \,
\rmd \Vector{W}_{\nu}
\label{eq:parsde}
\end{equation}
as a trajectory-wise approximation to $\br_{\nu}$. Here $\bm{\Vector{W}}_{\nu}$ is a Wiener process, and the $N \times N$ matrix $H_{\mu\nu}$ is the equilibrium average of the diffusion tensor $\Tensor{D}_{\mu\nu}$ (see Supporting Information), given by
\begin{equation}
H_{\mu\nu}  = \delta_{\mu\nu}  + (1 - \delta_{\mu\nu}) \, \bar H_{\mu\nu}
\label{eq:Hmunu}
\end{equation}
The expression for the matrix $\bar H_{\mu\nu}$ is discussed shortly below. The matrix $S_{\mu\nu}$ satisfies the expression,
\begin{equation}
\sum_{\alpha} S_{\mu\alpha}  \, S_{\nu\alpha} =  \bar H_{\mu\nu} \, ,     \quad \text{for} \quad \mu \ne \nu
\end{equation}
Note that, $\bar H_{\mu\mu} = S_{\mu\mu} =1$. The equilibrium average of $\Tensor{D}_{\mu\nu}$  is carried out with the equilibrium distribution function in the absence of excluded volume interactions, since an analytical solution for the distribution function is only known under $\theta$-solvent conditions. The advantage of using \cref{eq:parsde} for the purpose of variance reduction comes from the fact that Fixman has previously calculated $\bar H_{\mu\nu}$ and $\langle {\hat \CS (t)} \rangle_{\mathrm{eq}}$  analytically for the RPY tensor~\cite{Fixman1983,Fixman1981}. By simulating~\cref{eq:parsde} simultaneously with the stochastic differential equation for $\br_{\nu}$ (see Supporting Information), with the same Weiner process $\Vector{W}_{\nu}$, the fluctuations  ${\hat f}_{\text{C}_\text{S}}$ can be estimated, and consequently the mean value of the control variate, $\left\langle \hat E_{\text{C}_\text{S}}\right\rangle_{\mathrm{eq}}$. For the sake of completeness, we reproduce Fixman's expressions for $\bar H_{\mu\nu}$ and $\langle {\hat \CS (t)} \rangle_{\mathrm{eq}}$, with the non-dimensionalization scheme and notation used here, in the Supporting Information,

The efficacy of the variance reduction procedure is demonstrated in~\cref{fig:vred}, where the various auto-correlation functions obtained from the simulation of a bead-spring chain under $\theta$-conditions, with $N=18$, and $\hsH = 0.25$, are displayed.  The positive
correlation between the two functions \CS\ and ${\hat \CS}$, and the
reduction in the variance in $\hat E_{\text{C}_\text{S}}$ can be clearly observed.

Variance reduction was used here only for simulations with $z=0$
($\theta$-solvent), $z=0.01$, and $z=0.1$. For higher $z$, the correlation
between the two stochastic processes was lost and there was no benefit in using $\hat E_{\text{C}_\text{S}}$ in place of \CS. This is not unexpected since the equilibrium averaging of the diffusion tensor is carried out with the equilibrium distribution function in the absence of excluded volume interactions.

The stress-stress auto-correlation function must be integrated to obtain the intrinsic viscosity, as can be seen from~\cref{eq:etap0gk}, where, when appropriate, we use the  control variate $\hat E_{\text{C}_\text{S}}(t)$ instead of $\CS(t)$. In spite of the reduced variance, the numerical integration of this function is subject to errors. Consequently, we use a non-linear least square fit of the auto-correlation function instead, and evaluate the integral of the fitting function. Details are given in the Supporting Information.

\section{\label{sec:resultexpt} Results and Discussion}

\subsection{\label{subsec:ivisc} Intrinsic viscosity of DNA solutions}

The intrinsic viscosity of a polymer solution is typically obtained from a virial expansion of the dilute solution viscosity as a function of concentration. Two commonly used forms of the virial expansion are the Huggins equation,
\begin{equation}
  \eta_\text{sp} \equiv \frac{\eta_{\mathrm{p, 0}}}{\etas} =  \ivisc\
  c + \kh \left(\ivisc \, c\right)^{2} + \kh^{\prime} \left(\ivisc \,
    c\right)^{3} + \cdots
\label{eq:Hug}
\end{equation}
and Kraemer's equation,
\begin{equation}
\label{eq:Kra}
\ln\frac{\etao}{\etas} = \ivisc\ c - \kk \left(\ivisc \, c\right)^{2}
+ \kk^{\prime} \left(\ivisc \, c\right)^{3} + \cdots
\end{equation}
where, $\eta_\text{sp}$ is the specific viscosity, the coefficient \kh\ in the quadratic term in Huggins equation~(\cref{eq:Hug}) is the Huggins constant, and is analogous to the second virial coefficient for viscosity~\cite{RubCol03}, while \kk\ is the equivalent coefficient in Kraemer's equation. The parameters $\kh^{\prime}$ and $\kk^{\prime}$ are coefficients of the cubic terms in the Huggins and Kraemer's equations, respectively.

Substituting the Huggins expansion in terms of $\etao$ from~\cref{eq:Hug} into the left hand side of Kraemer's equation~(\cref{eq:Kra}), and comparing terms of similar order leads to,
\begin{equation}
\label{eq:Kracoeff}
\kk = \frac{1}{2} - \kh \, , \quad \text{and} \quad \kk^{\prime} = \kh^{\prime} - \kh + \frac{1}{3}
\end{equation}
Typically, dilute solution viscosities are measured at low values of concentration, where the contribution of the cubic term in the Huggins equation is negligible. As a result, by plotting $\eta_\text{sp}/c$ versus concentration, the intrinsic viscosity can be obtained from the intercept on the $y$-axis of a straight line fitted to the data, while \kh\ can be determined from the slope of the line, since,
\begin{equation}
\label{eq:huggins}
\frac{\eta_\text{sp}}{c } = \ivisc\ + \kh\ \ivisc^{2} c
\end{equation}
As pointed out by \citet{Pamies20081223} even though $\kh^{\prime} \left(\ivisc \, c\right)^{3} \approx 0$, the contribution of the cubic term in Kraemer's equation need not be zero (unless, $\kh \approx 1/3$, see~\cref{eq:Kracoeff}). At sufficiently low concentrations, however,  Kraemer's equation~(\cref{eq:Kra}) suggests that $[\ln (\eta_{0}/ \etas)]/ c$ will be linear in concentration,
\begin{equation}
\label{eq:kraemer}
\frac{1}{c} \, \ln\frac{\etao}{\etas} = \ivisc\  - \kk\ \ivisc^{2} c
\end{equation}
As a result, the intrinsic viscosity can be obtained from the intercept of a line fitted to measurements of $[\ln (\eta_{0}/ \etas)]/ c$ versus $c$ (in a so-called Fuoss-Mead plot~\cite{Mead42}), while \kk\ can be determined from the slope of the line.

Since the leading order term in the expansions for both $\eta_\text{sp}$ and $\ln ({\etao}/{\etas})$ is $\ivisc c$, \citet{Solomon1962683} suggested that the virial expansion of the difference $\eta_\text{sp} - \ln ({\etao}/{\etas})$ would have a weaker dependence on concentration,
\begin{equation}
\label{eq:SC}
\eta_\text{sp} - \ln\frac{\etao}{\etas} = \ksc \left(\ivisc \, c\right)^{2} + \ksc^{\prime} \left(\ivisc \, c\right)^{3} + \cdots
\end{equation}
\begin{equation}
\label{eq:SCcoeff}
\text{with}, \quad \ksc = \frac{1}{2} \, , \quad \text{and}  \quad  \ksc^{\prime} = \kh - \frac{1}{3}
\end{equation}
As a result, by defining the quantity,
\begin{equation}
\label{eq:etac}
[\eta]_{\text{c}} = \frac{1}{c} \sqrt{2   \left( \eta_\text{sp}  - \ln\left({\etao}/{\etas}\right) \right)}
\end{equation}
it follows that,
\begin{equation}
\label{eq:etasc}
[\eta]_{\text{c}} = \ivisc +  \ksc^{\prime} \ivisc^{2} \, c + \cdots
\end{equation}
As discussed in some detail by \citet{Pamies20081223}, under the special circumstances when $\ksc^{\prime} \ivisc^{2} \, c \approx 0$, or $\kh \approx 1/3$ (see~\cref{eq:SCcoeff}), the intrinsic viscosity can be determined from the Solomon-Ciut\u{a} equation~(\cref{eq:etasc}) by measuring the viscosity at a single concentration, without the necessity of an extrapolation procedure. The departure of $[\eta]_{\text{c}}$ from a constant value when $[\eta]_{\text{c}}$ is plotted as a function of $c$, can be seen as indicating the departure of \kh\ from a value of 1/3.

\begin{figure*}[tbp]
\begin{center}
\begin{tabular}{cc}
\resizebox{0.39\linewidth}{!} {\includegraphics*{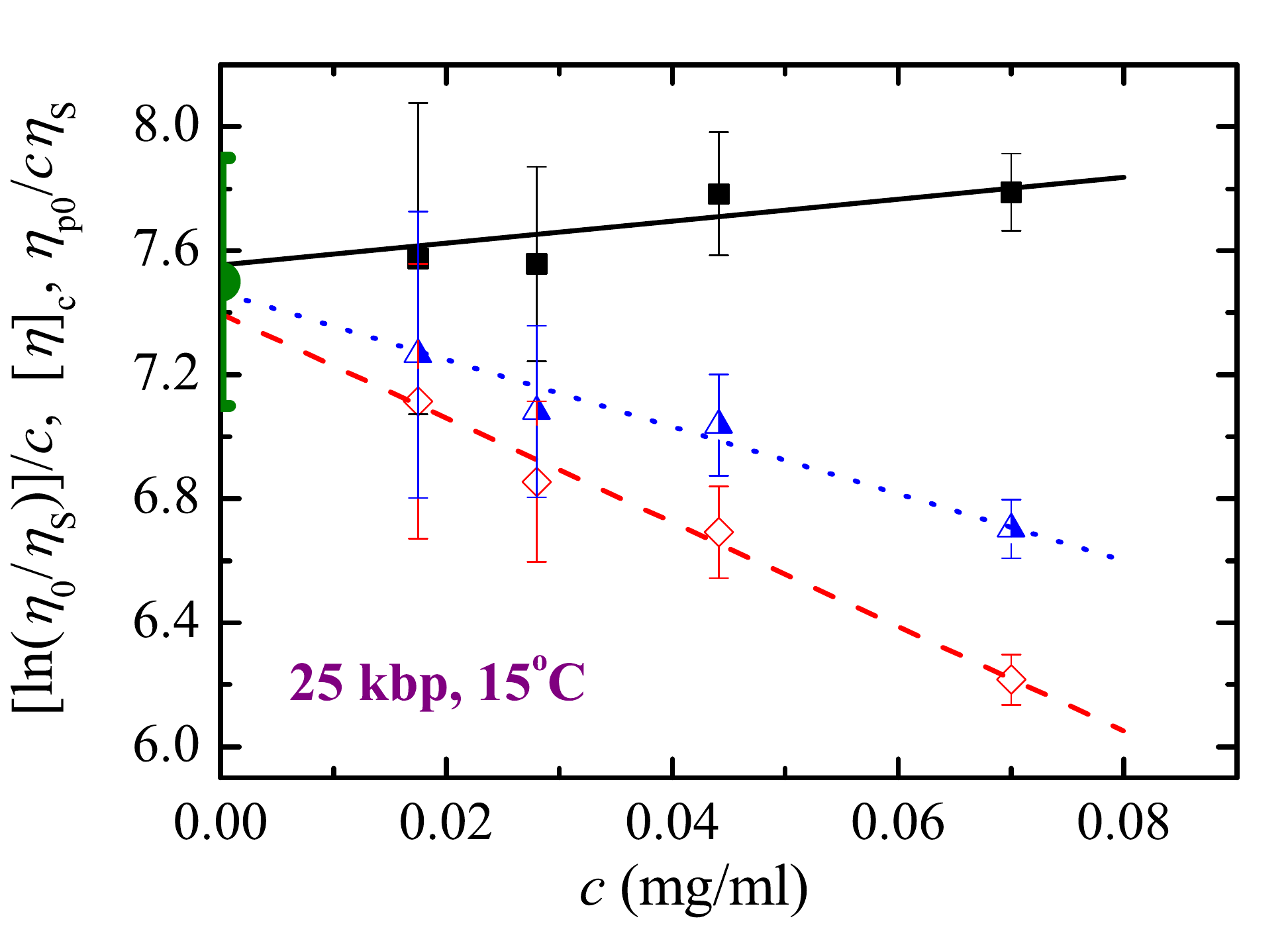}}&
\resizebox{0.39\linewidth}{!} {\includegraphics*{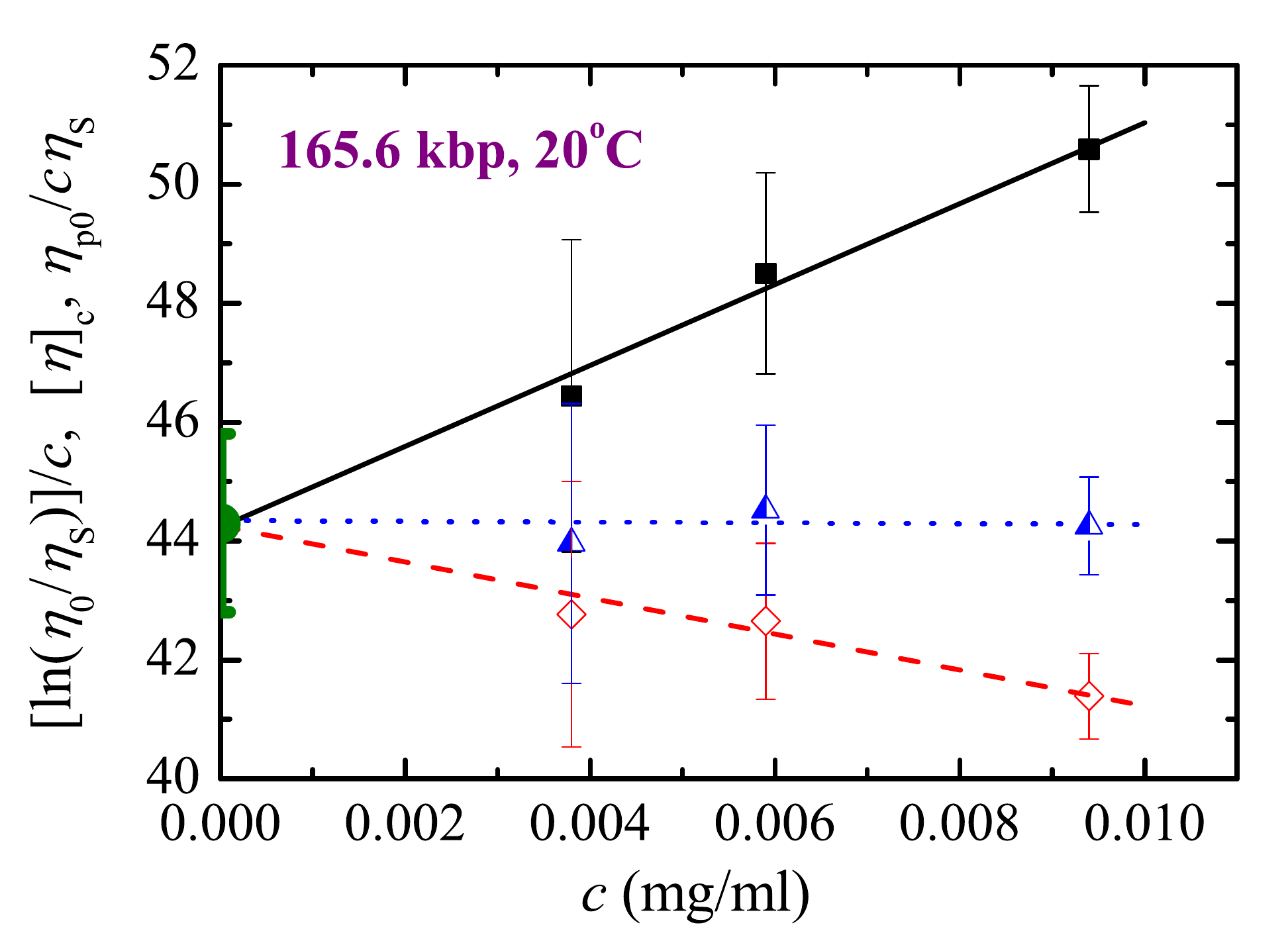}}\\
(a) & (b)  \\
\resizebox{0.39\linewidth}{!} {\includegraphics*{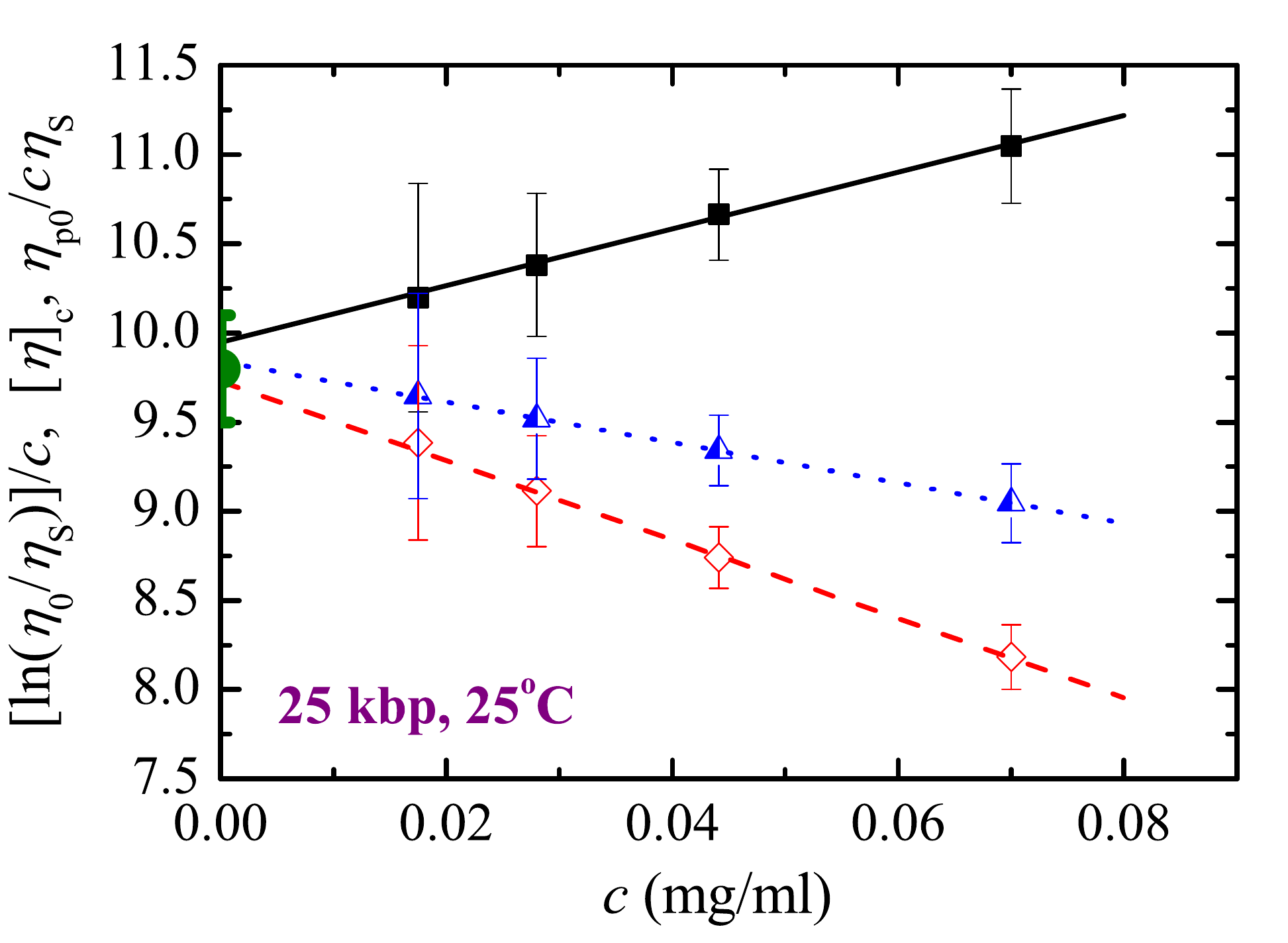}}&
\resizebox{0.39\linewidth}{!} {\includegraphics*{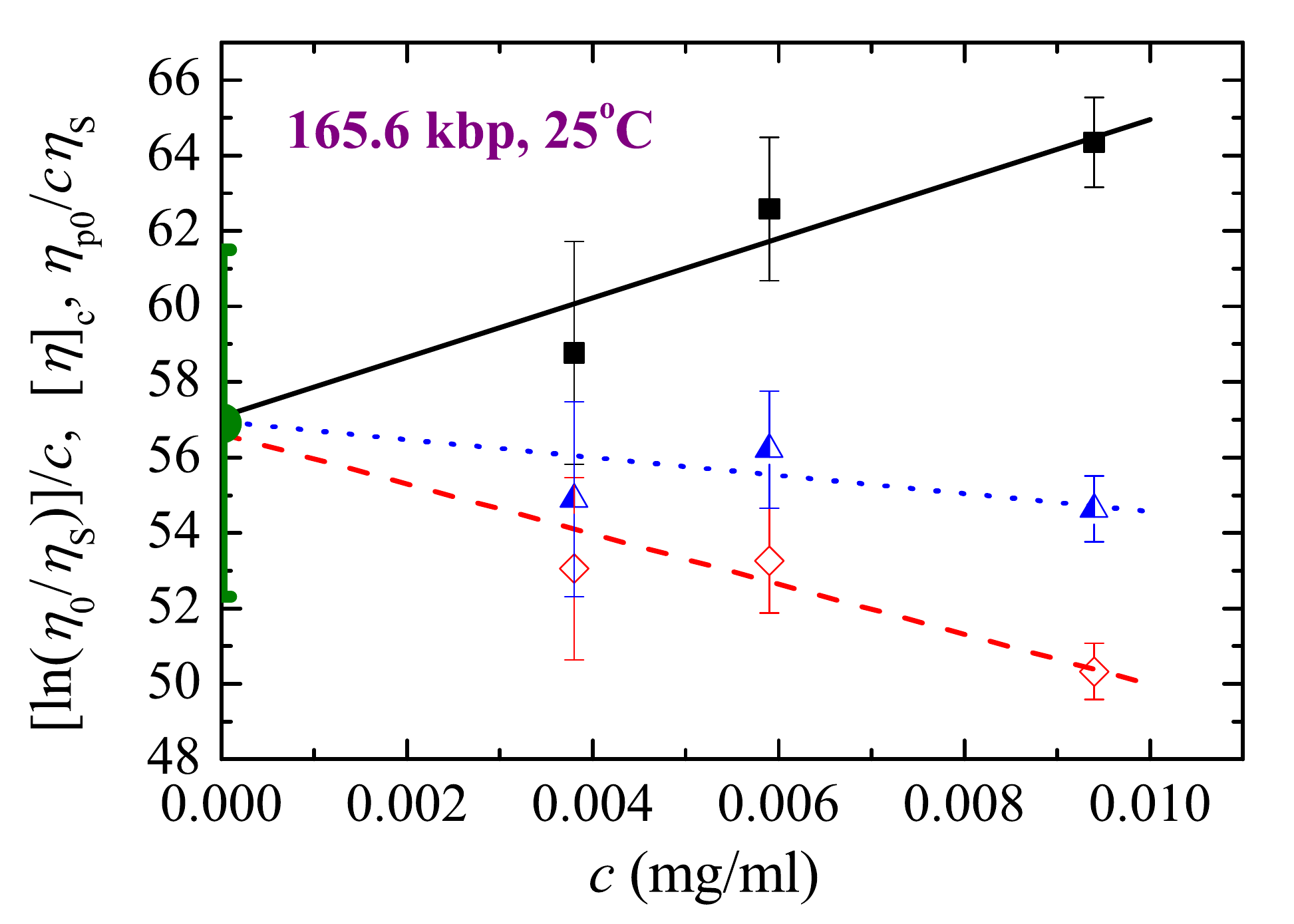}}\\
(c) & (d)  \\
\resizebox{0.39\linewidth}{!} {\includegraphics*{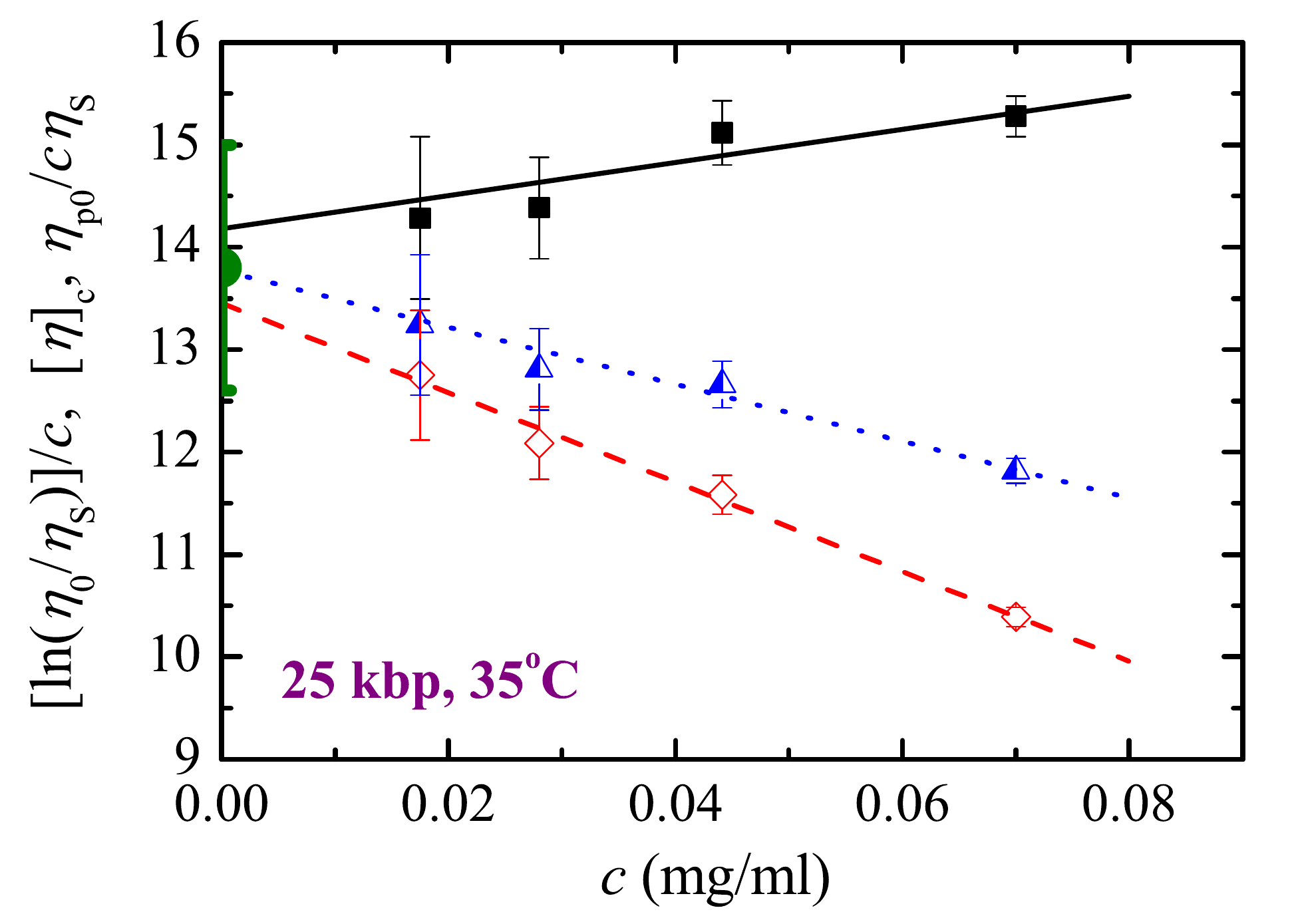}}&
\resizebox{0.39\linewidth}{!} {\includegraphics*{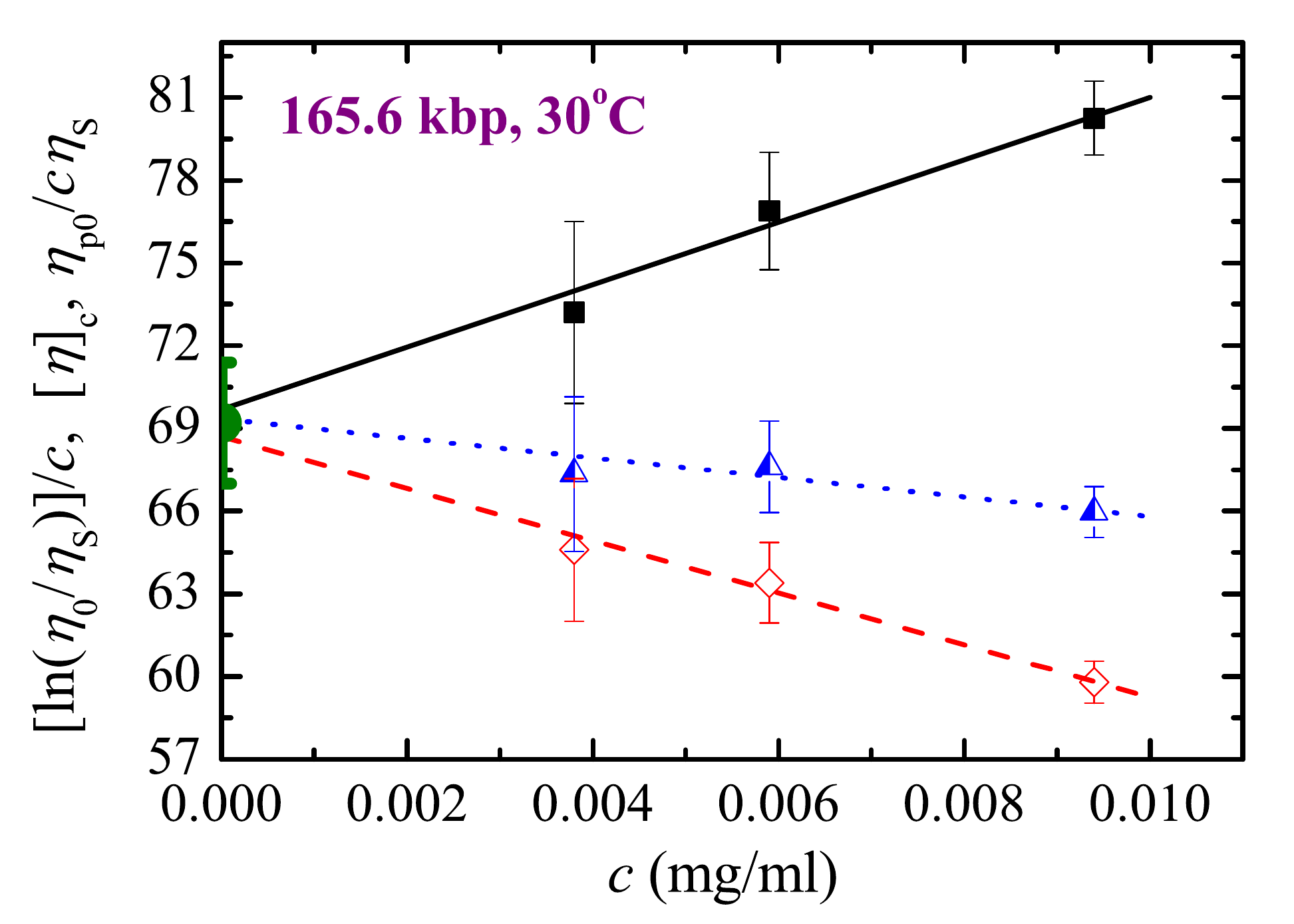}}\\
(e) & (f)  \\
\end{tabular}
\end{center}
\vskip-10pt
\caption{\label{fig:ivisc} Determination of \ivisc\ for 25 kbp and T4 DNA. The left and right column of figures represent 25 kbp and T4 DNA respectively at different temperatures (indicated within the figures). The solid, dashed and dotted lines are least-squares linear fits to the data points extrapolated to zero concentration in accordance with the Huggins, Kraemer and Solomon-Ciut\u{a} equations, respectively. In each figure, the mean value of \ivisc\ obtained by extrapolating data for $[\ln (\eta_{0}/ \etas)]/ c$ (open diamonds), $\eta_{\mathrm{p, 0}} / c \etas$ (filled squares) and \etac\ (half-filled triangles) to zero concentration, is represented by an filled circle (the common intercept on the $y$-axis).  Note that the quantities on the $y$-axis are in units of ml/mg, the same as \ivisc.}
\end{figure*}

\begin{figure}[tbp]
\begin{center}
\resizebox{0.9\linewidth}{!} {\includegraphics*{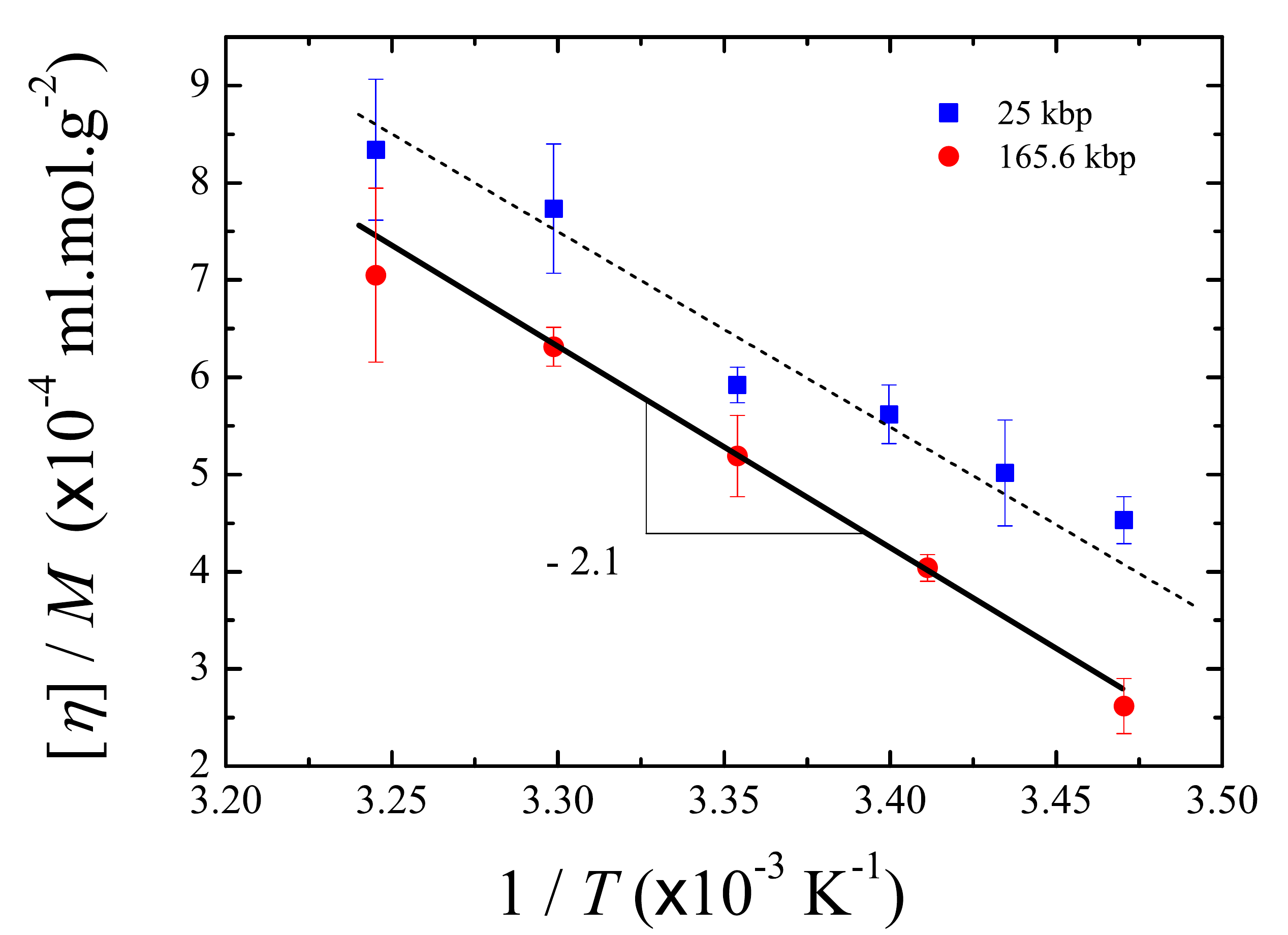}}
\end{center}
\caption{\label{fig:iviscvsT} Temperature dependence of $(\ivisc/M)$ for  25 kbp DNA and T4 DNA. The line through the T4 data is a least-squares linear fit, while the line through the 25 kbp data, which is more scattered, is drawn with the same slope to guide the eye.}
\end{figure}

Plots of the relevant variables in the linear versions of the Huggins
equation (\cref{eq:huggins}), the Kraemer equation (\cref{eq:kraemer})
and the Solomon-Ciut\u{a} equation (\cref{eq:etasc}), as a function of
concentration, can now be interpreted in the light of the discussion
above. \cref{fig:ivisc} displays plots of ${\eta_\text{sp}}/{c }$, $[\ln (\eta_{0}/ \etas)]/ c$, and $[\eta]_{\text{c}}$, obtained using results of
the zero shear rate solution viscosity measurements, as a function of
concentration. Values of \ivisc\ obtained by extrapolating linear fits
to the finite concentration data to the limit of zero concentration
are listed in~\cref{tab:ivisc}, where the subscript on \ivisc\
indicates the equation used to obtain the value. The mean values of
\ivisc\ obtained from the three methods are also indicated in the
table. It is clear that the three extrapolation methods give values
that are fairly close to each other.

Recently, \citet{Rushing2003} have shown that, in line with a relationship proposed originally by \citet{Stockmayer1963}, the ratio $(\ivisc/M)$ for a number of different polymer-solvent systems scales linearly with inverse temperature, with a slope that is independent of molecular weight. \cref{fig:iviscvsT} indicates that the mean value of $(\ivisc/M)$, for both the DNA samples, scales linearly with inverse temperature as $T$ increases from $T_\theta$ to good solvent conditions, with a slope that is common  for both the DNA, in agreement with the observations of \citet{Rushing2003} for synthetic polymer solutions.

As discussed earlier, the values of \kh\ can be obtained from the slopes of the lines in~\cref{fig:ivisc}. While it is obtained directly from the slope of the line through the Huggins data, Kraemer's data gives \kh\ from \kk\ [see \cref{eq:Kracoeff}], and the Solomon-Ciut\u{a} data gives \kh\ from $\ksc^{\prime}$  [see \cref{eq:SCcoeff}]. The values of \kh\ obtained from these different methods are listed in~\cref{tab:constants}. We first discuss the data for T4 DNA, which appears to be more in line with previous observations on synthetic polymer solutions.

\citet{Pamies20081223} have recently tabulated values of \kh\ for several systems by collating data reported previously in literature (see Table~1 in Ref.~\citenum{Pamies20081223}). For flexible polymers, \kh\ is observed to lie in the range $0.4-0.7$ for $\theta$-solvents, and in the range $0.2-0.4$ for good solvents. Clearly, values of \kh\ reported for T4 DNA in~\cref{tab:constants} lie in the expected ranges for $\theta$ and good solvents, with the $\theta$-solvent value greater than that for good solvents. The three different means of estimating \kh\ also give values reasonably close to each other. Since $\kh \approx 1/3$, we expect from the Solomon-Ciut\u{a} equation~(\cref{eq:etasc}) that the slope of the line through measured values of \etac\  as a function of concentration should be close to zero. This is indeed the case, as can be seen from Figs.~(b), (d) and (f) for T4 DNA in~\cref{fig:ivisc}.

\begin{figure}[tbp]
\begin{center}
\begin{tabular}{c}
\resizebox{0.8\linewidth}{!} {\includegraphics*{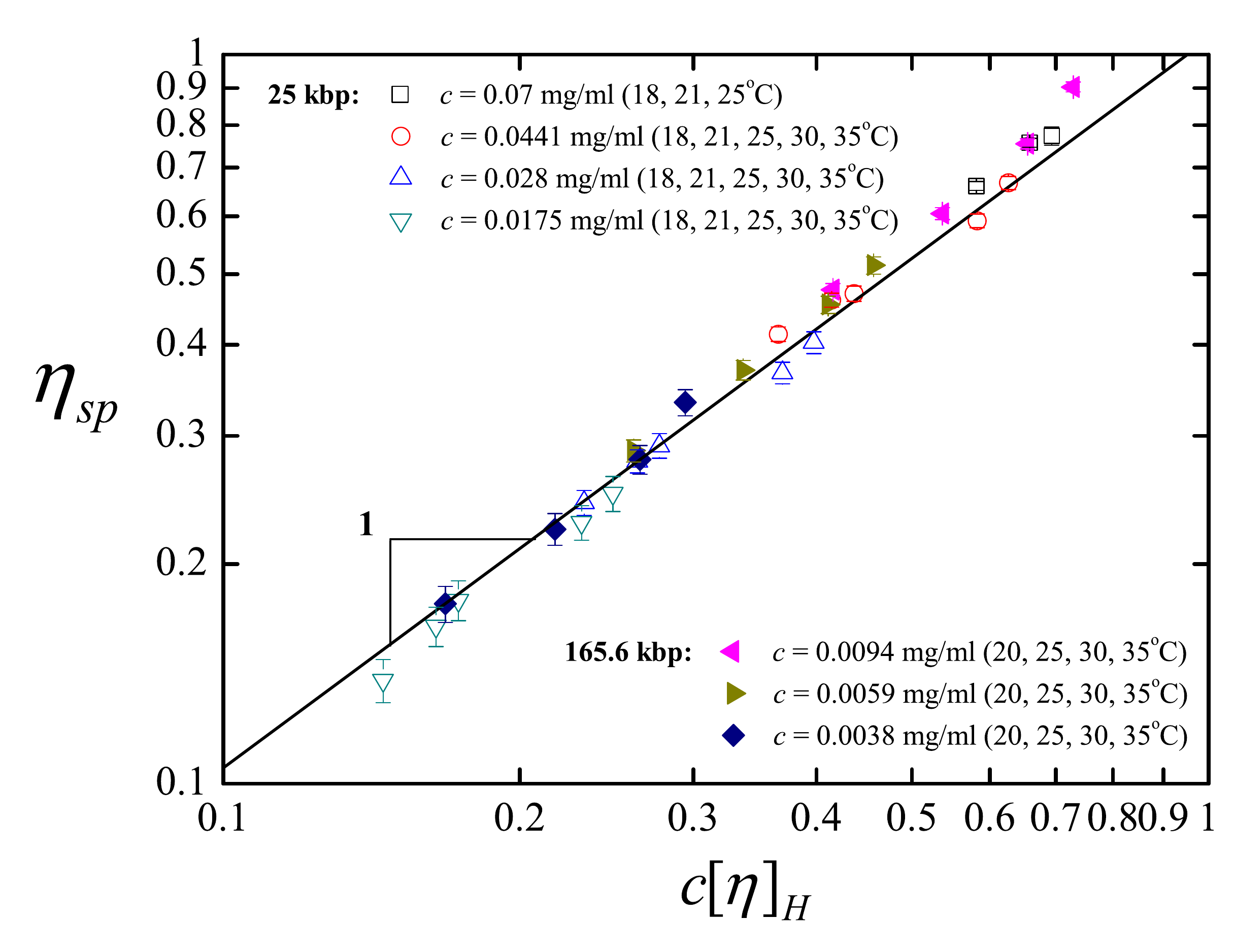}} \\
(a)  \\
\resizebox{0.8\linewidth}{!} {\includegraphics*{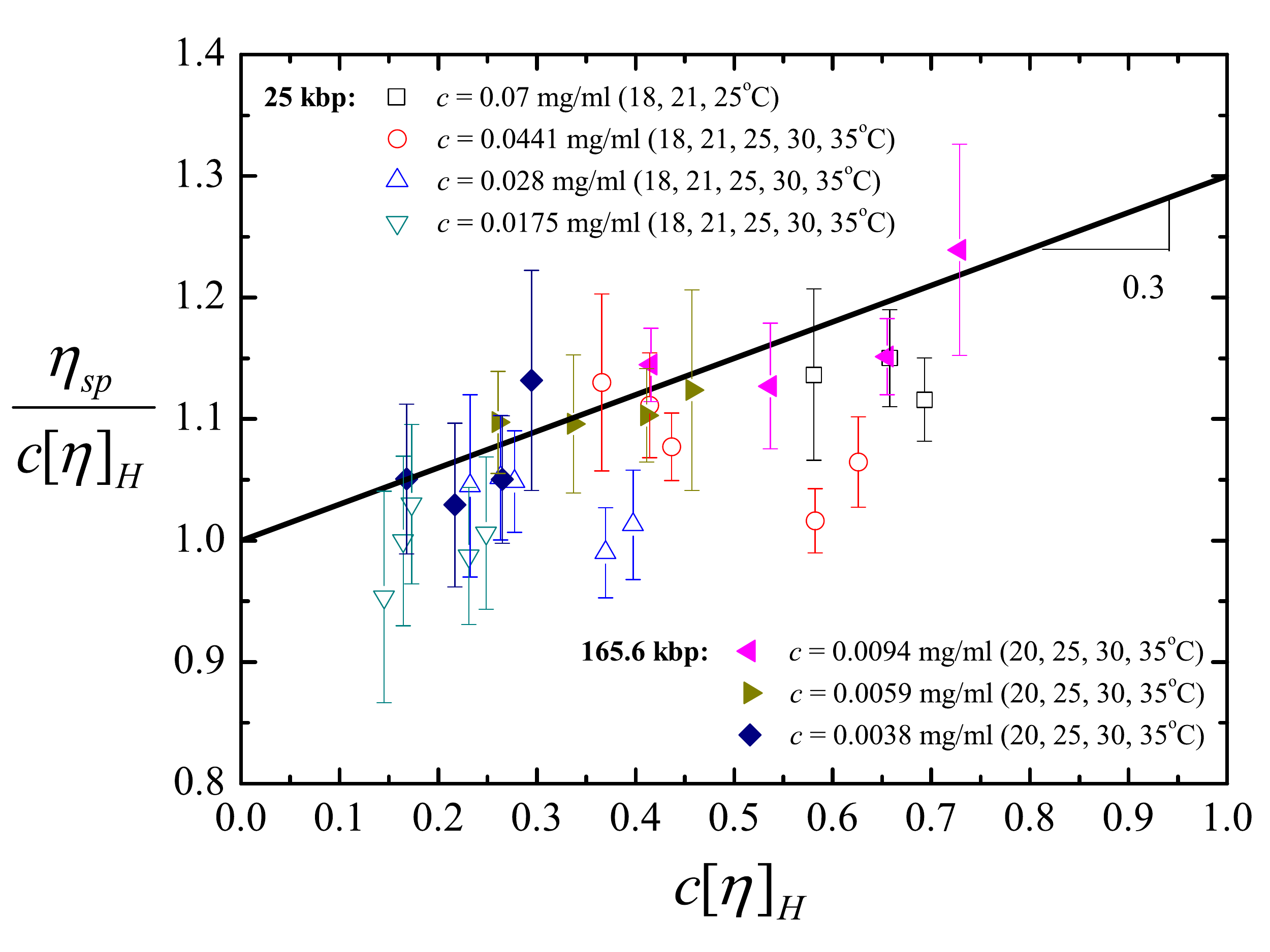}} \\
(b)  \\
\end{tabular}
\end{center}
\caption{(a) Dependence of the specific viscosity $\etasp$ on the non-dimensional concentration $c [\eta]_{\mathrm{H}}$, and, (b) dependence of the dimensionless ratio $\etasp / c [\eta]_{\mathrm{H}}$ on $c [\eta]_{\mathrm{H}}$, for the two DNA used in this work at different absolute concentrations, each of which is at different temperatures in good solvents.}
\label{fig:dilutehuggins}
\end{figure}

When the term of order $(\ivisc c)^{3}$ is negligible, we expect a plot of $\eta_\text{sp}$ versus $c \ivisc $ to depend quadratically on $c \ivisc $ for increasing values of $c \ivisc $ (see~\cref{eq:Hug}).  The departure from linearity can  be observed for the T4 DNA data in~\cref{fig:dilutehuggins}~(a) for $c \ivisc \gtrsim 0.3$ (filled symbols). The importance of the quadratic term can be seen more clearly by plotting $\eta_\text{sp}/ (c \ivisc ) $ versus $(c \ivisc )$, as shown

\begin{landscape}
\begin{table*}[!t]
\addtolength{\tabcolsep}{2pt}
\caption{\label{tab:ivisc} Intrinsic viscosities [$\eta$] (in ml/mg) for 25 kbp and T4 DNA at various temperatures ($T$), as obtained from different extrapolation methods: Huggins ($[\eta]_{\mathrm{H}}$), Kraemer ($[\eta]_{\mathrm{K}}$) and Solomon-Ciut\u{a} ($[\eta]_{\mathrm{SC}}$). The mean of the [$\eta$] values from these extrapolations are also indicated at each temperature. The swelling ratio \aeta\ is also listed for each DNA at each temperature and has been calculated based on the $[\eta]_{\mathrm{mean}}$ values. Note that $T_{\theta}$ = 15 \degC.}
\vspace{2pt}
\centering
\begin{small}
\begin{tabular}{l l l l l l| l l l l l}
\hline
$T$ & \multicolumn{5}{c}{25 kbp} & \multicolumn{5}{c}{T4 DNA} \\
\cline{2-6} \cline{7-11}
 (\degC) & $[\eta]_{\mathrm{H}}$ & $[\eta]_{\mathrm{K}}$ & $[\eta]_{\mathrm{SC}}$ & $[\eta]_{\mathrm{mean}}$ & \aeta\ & $[\eta]_{\mathrm{H}}$ & $[\eta]_{\mathrm{K}}$ & $[\eta]_{\mathrm{SC}}$ & $[\eta]_{\mathrm{mean}}$ & \aeta\ \\ [0.5ex]
\hline
\hline
15  & 7.6 $\pm$ 0.1 & 7.4 $\pm$ 0.1 & 7.5 $\pm$ 0.1 & 7.5 $\pm$ 0.4 & 1 $\pm$ 0.03 & 28.5 $\pm$ 1.4 & 28.9 $\pm$ 1.3 & 28.8 $\pm$ 1.3 & 28.7 $\pm$ 3.1 & 1 $\pm$ 0.05  \\

18 & 8.3 $\pm$ 0.5  & 8.3 $\pm$ 0.4 & 8.4 $\pm$ 0.4 & 8.3 $\pm$ 0.9 & 1.03 $\pm$ 0.04 & -- & -- & -- & -- & -- \\

20 & --  & -- & -- & -- & -- & 44.2 $\pm$ 0.7  & 44.3 $\pm$ 0.6  & 44.3 $\pm$ 0.6  & 44.3 $\pm$ 1.5 & 1.15 $\pm$ 0.04  \\

21 & 9.4 $\pm$ 0.3  & 9.3 $\pm$ 0.2 & 9.3 $\pm$ 0.2 & 9.3 $\pm$ 0.5 & 1.07 $\pm$ 0.03 & -- & -- & -- & -- & -- \\

25 & 9.9 $\pm$ 0.1  & 9.7 $\pm$ 0.1 & 9.8 $\pm$ 0.1 & 9.8 $\pm$ 0.3 & 1.09 $\pm$ 0.02 & 57.1 $\pm$ 2.4  & 56.6 $\pm$ 1.7  & 57 $\pm$ 2 & 56.9 $\pm$ 4.6 & 1.26 $\pm$ 0.06  \\

30 & 13.2 $\pm$ 0.2 & 12.4 $\pm$ 0.1 & 12.7 $\pm$ 0.1 & 12.8 $\pm$ 1.1 & 1.19 $\pm$ 0.04 & 69.7 $\pm$ 1.5  & 68.7 $\pm$ 0.8  & 69.3 $\pm$ 1.1  & 69.2 $\pm$ 2.2 & 1.34 $\pm$ 0.05 \\

35 & 14.2 $\pm$ 0.4 & 13.5 $\pm$ 0.1 & 13.8 $\pm$ 0.2 & 13.8 $\pm$ 1.2 & 1.22 $\pm$ 0.04 & 77.5 $\pm$ 5.3  & 76.8 $\pm$ 3.7  & 77.5 $\pm$ 4.1  & 77.3 $\pm$ 9.8 & 1.39 $\pm$ 0.08 \\
\hline
\end{tabular}
\end{small}
\end{table*}
\begin{table*}[t]
\begin{small}
\addtolength{\tabcolsep}{2pt}
\caption{\label{tab:constants} \kh\ as obtained from Huggins, Kraemer and Solomon-Ciut\u{a} equations for 25 and T4 DNA at different temperatures.}
\vspace{2pt}
\centering
\begin{tabular}{ccccccc}
\hline
$T$ (\degC) & \multicolumn{2}{c}{\kh\ (Huggins)} & \multicolumn{2}{c}{\kh\ (From Kraemer, see \cref{eq:Kracoeff})} & \multicolumn{2}{c}{\kh\ (From Solomon-Ciut\u{a}, see \cref{eq:SCcoeff})}\\
\cline{2-3} \cline{4-5} \cline{6-7}
 & 25 kbp & T4 DNA  & 25 kbp & T4 DNA  & 25 kbp & T4 DNA \\
\hline
\hline
15 ($T_{\theta}$) & 0.06 $\pm$ 0.04 & 0.82 $\pm$ 0.22   & 0.19 $\pm$ 0.02  & 0.64 $\pm$ 0.18   & 0.14 $\pm$ 0.3  & 0.68 $\pm$ 0.19   \\
18 & 0.24 $\pm$ 0.13 & --                & 0.28 $\pm$ 0.09  & --                & 0.25 $\pm$ 0.1  & --                \\
20 & --              & 0.35 $\pm$ 0.05   & --               & 0.35 $\pm$ 0.03   & --              & 0.33 $\pm$ 0.04   \\
21 & 0.24 $\pm$ 0.05 & --                & 0.3 $\pm$ 0.03   & --                & 0.26 $\pm$ 0.04 & --                \\
25 & 0.16 $\pm$ 0.01 & 0.24 $\pm$ 0.09   & 0.27 $\pm$ 0.01  & 0.29 $\pm$ 0.06   & 0.21 $\pm$ 0.01 & 0.26 $\pm$ 0.07   \\
30 & 0.01 $\pm$ 0.02 & 0.23 $\pm$ 0.04   & 0.22 $\pm$ 0.01  & 0.3 $\pm$ 0.02    & 0.14 $\pm$ 0.01 & 0.26 $\pm$ 0.03   \\
35 & 0.08 $\pm$ 0.03 & 0.32 $\pm$ 0.12   & 0.26 $\pm$ 0.01  & 0.35 $\pm$ 0.08   & 0.18 $\pm$ 0.02 & 0.31 $\pm$ 0.08   \\
\hline
\end{tabular}
\end{small}
\end{table*}
\end{landscape}

\noindent in~\cref{fig:dilutehuggins}~(b), since,
\begin{equation}
\label{eq:huggins2}
\frac{\eta_\text{sp}}{c \ivisc} = 1 + \kh\ c \ivisc
\end{equation}
The data for T4 DNA is scattered around a line with slope = 1/3, as expected from the values of \kh\ listed for T4 DNA in~\cref{tab:constants}.

Values of \kh\ extracted from the dilute solution viscosity data for 25 kbp DNA using the Huggin's method have a greater degree of uncertainty associated with them compared to those for T4 DNA (see first column in~\cref{tab:constants}). Even though the values obtained from the Kraemer and Solomon-Ciut\u{a} equations lie closer to the expected range of values for good solvents, the $\theta$-solvent values are smaller than the good solvent values. \cref{fig:dilutehuggins}~(a) indicates that the dependence of ${\eta_\text{sp}}$ on $c [\eta]_{\mathrm{H}}$ for 25 kbp DNA appears to be linear in the entire range of values of $c [\eta]_{\mathrm{H}}$ observed here (empty symbols), which suggests that it would be harder to extract the values of \kh\ with confidence using the Huggin's method. This is also clearly reflected in~\cref{fig:dilutehuggins}~(b), where the data indicates that the value of the Huggins constant is highly scattered, and in most cases smaller than 1/3. More extensive measurements at a larger range of concentrations would be required to obtain \kh\ with greater accuracy for 25 kbp DNA.

The intrinsic viscosity data obtained at various temperatures can be used to calculate the viscosity radius of 25 kbp and T4 DNA. Of the two properties of interest in the present work, namely, $U_{\eta R}$ and $\alpha_{\eta}$, the latter is directly calculable from experimental measurements. Values for the two DNA samples are reported in~\cref{tab:ivisc}. On the other hand, the direct estimation of  \Uer\  requires the additional knowledge of \Rg.  While the prediction of  \Uer\ here by simulations is based on the determination of both the viscosity and the radius of gyration as a function of solvent quality, we do not have experimental information on \Rg\ for the two DNA samples studied here.  However, it is clear from \cref{eq:uetaratio1} that the ratio $(U_{\eta R}/U_{\eta R}^{\theta})$  can be calculated without a knowledge of \Rg, if  the dependence of \aeta\ and \ag\ on solvent quality is known.

In the context of determining the dependence of \ah\ on solvent quality for DNA, \citet{Pan2014339} established the relationship between pairs of values of $T$ and $M$, and $z$, assuming that DNA is a flexible molecule at the molecular weights that were considered. Here, we take into account the wormlike nature of DNA molecules, and show in the Supporting Information, that a mapping between $T$ and $M$ and the parameter $\tilde z$ can be constructed, similarly. As a result, since the swelling $\alpha_{\eta}$ is known for the two DNA samples at various values of $T$ (\cref{tab:ivisc}), we can determine the dependence of \aeta\ on $\tilde z$ for these two samples. The determination of the dependence of \ag\ on $\tilde z$ is discussed below.

As mentioned earlier, the quasi-two-parameter theory is an extension of the two-parameter theory to account for chain stiffness~\cite{yamakawa1997}. Essentially, the theory assumes that functional forms of universal crossover functions for wormlike chains are \emph{identical} to those for flexible chains, with the excluded volume parameter $z$ replaced by the parameter $ \tilde z$. As a consequence, the quasi-two-parameter theory expects the Domb-Barrett and Barrett equations for \ag\ and \aeta, respectively, to successfully describe the swelling of the radius of gyration and the viscosity radius of wormlike chains, when $z$ is replaced by $\tilde z$. This expectation has been shown to be exceedingly well fulfilled for a range of experimental data for a variety of polymer-solvent systems~\cite{Osa2001,Tominaga20021381}. Here, we assume analogously that the functional form used to fit BD data for the swelling of the radius of gyration of flexible chains, can be used to describe the swelling of wormlike chains, by replacing $z$ with $\tilde z$. As a result, the dependence of \ag\ on $\tilde z$ can be obtained from~\cref{eq:ag}, and the experimentally measured dependence of $(U_{\eta R}/U_{\eta R}^{\theta})$ on $\tilde z$ can be determined from \cref{eq:uetaratio1}, using experimentally measured values of \aeta, and BD simulation results for \ag.

The procedure outlined above enables a comparison of experimentally measured values of $\alpha_{\eta}$ and $(U_{\eta R}/U_{\eta R}^{\theta})$ for DNA, at identical values of the solvent quality $\tilde z$, with earlier observations for synthetic polymer solutions and with results of Brownian dynamics simulations, as discussed in the following sections.

\subsection{\label{subsec:uetartheta} Universal viscosity ratio under $\theta$-conditions}

The zero shear rate viscosity, in the absence of hydrodynamic interactions, is related to the radius of gyration by the following expression,
\begin{equation} \label{etarg}
\eta^{*}_{\mathrm{p, 0}} =  \frac{2}{3} \, N \, {\Rg^{*}}^2
\end{equation}
which can be derived by developing a retarded motion expansion for the stress tensor~\cite{prakash:macromol-01}. As a result, $\eta_{\mathrm{p, 0}}$ scales with $N$ as $N^{2}$, and in the absence of  hydrodynamic interactions, the ratio $U_{\eta R}^{\theta}$  is not a universal constant since it scales with $N$ as $N^{1/2}$  (see \cref{eq:uetarsim}). It becomes a universal constant only when hydrodynamic interactions are included in the model since this alters the scaling of $\eta_{\mathrm{p, 0}}$ with $N$ from $N^{2}$ to $N^{3/2}$, as first demonstrated by Zimm theory~\cite{RubCol03} and by two-parameter theories which include pre-averaged hydrodynamic interactions~\cite{Barrett1984}.

\begin{figure}[!htbp]
\begin{center}
\resizebox{0.9\linewidth}{!} {\includegraphics*{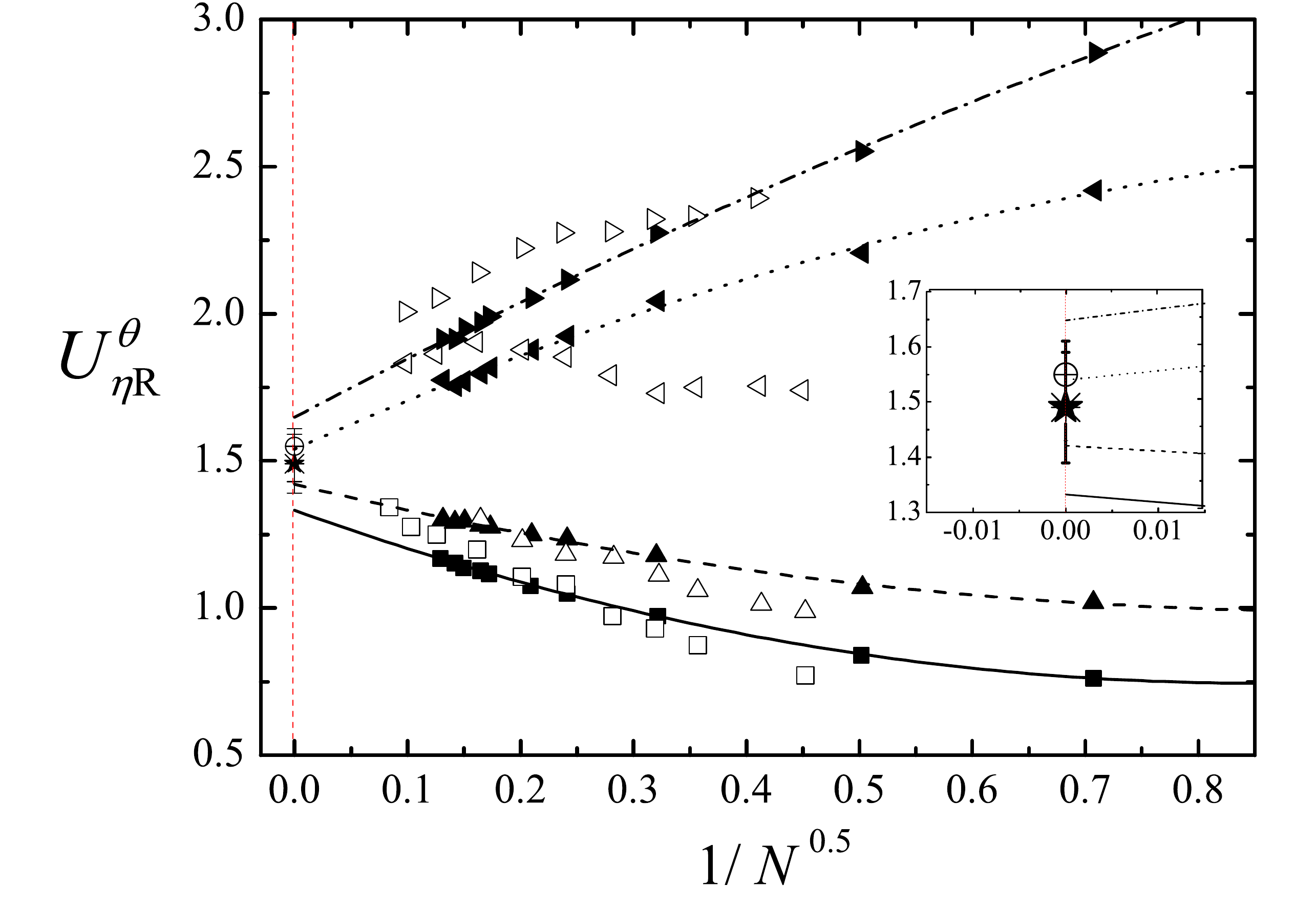}}
\end{center}
\vskip-20pt
\caption{Universal viscosity ratio for a $\theta$-solvent ($U_{\eta R}^{\theta}$). The filled symbols are the results of current BD
  simulations  determined using the Green-Kubo expression for the
  zero shear rate viscosity: $\blacksquare$  \hsH\ = 0.2 $\blacktriangle$ \hsH\ = 0.25 $\blacktriangleleft$ \hsH\  = 0.45 $\blacktriangleright$ \hsH\ = 0.5. The empty symbols are the results of non-equilibrium simulations at finite shear rate reproduced from \citet{Kroger20004767}: $\square$ \hsH\ = 0.2 $\triangle$ \hsH\ = 0.25 $\lhd$ \hsH\ = 0.45 $\rhd$ \hsH\ = 0.5. The solid (\hsH\ = 0.2), dashed (\hsH\ = 0.25), dotted (\hsH\ = 0.4) and dash-dotted (\hsH\ = 0.45) lines are second order polynomial fits to the
  current simulations data. The inset shows estimated asymptotic values of $U_{\eta R}^{\theta}$ (on the y-axis): current work ($\bigstar =1.49 \pm 0.1$), \citet{Kroger20004767} ($\oplus = 1.55 \pm 0.04$), and  \citet{Miyaki1980} ( $ \ast = 1.49 \pm 0.06$).}
\label{fig:uetartheta}
\end{figure}

The framework for getting universal predictions within the context of  Brownian dynamics simulations that include fluctuating hydrodynamic interactions has been clearly delineated by \citet{Kroger20004767}, who show that model independent predictions of several properties can be obtained by careful extrapolation of data accumulated for finite chains to the long chain limit. By carrying out non-equilibrium Brownian dynamics simulations at finite shear rates, and by extrapolating the finite shear rate data to the limit of zero shear rate, they have obtained equilibrium predictions of several properties. In particular, they predict $U_{\eta R}^{\theta} \approx1.55 \pm 0.04$.
In contrast to their approach, we have used a Green-Kubo expression (\cref{eq:etap0gk}) coupled with a variance reduction scheme in order to obtain predictions of the zero shear rate viscosity under $\theta$-solvent conditions. Results for $U_{\eta R}^{\theta}$ obtained by following this procedure are displayed in~\cref{fig:uetartheta}, where data at constant  \hsH, at several different chain lengths $N$, is extrapolated to $\Ntoinf$, which corresponds to the non-draining limit. The choice of ${1/\sqrt{N}}$ as the $x$-axis is made because the leading order correction to the infinite chain length limit value of universal ratios has been shown to be \OrderOf{1/\sqrt{N}} in Zimm theory~\cite{Osa72,ott87d},
and in simulations~\cite{Kroger20004767}. As is well known~\cite{ott87d,ottrab89}, there is a special value of $\hsH$ called the fixed point, denoted by $\hsf$, at which the leading order correction to the limiting value changes from being of \OrderOf{1/\sqrt{N}} to \OrderOf{1/{N}}, resulting in the asymptotic value being attained for smaller values of $N$. For pre-averaged hydrodynamic interactions, it is known that $\hsf=0.2424\ldots$~\cite{Osa72,ott87d}. It is also known that calculations of universal properties for values of $\hsH$ above and below $\hsf$, approach the long chain limit value along curves with slopes of opposite sign with increasing values of $N$. The choice of values of $\hsH$ in the current simulations have been motivated by these considerations, in order to obtain better estimates of long chain limit predictions. As can be seen from~\cref{fig:uetartheta}, values of $U_{\eta R}^{\theta}$ for $\hsH = 0.2$ and $\hsH = 0.25$  approach the long chain limit along curves whose slopes are of opposite sign to those for $\hsH = 0.45 $ and $ \hsH = 0.5$. This suggests that for simulations predictions of $U_{\eta R}^{\theta}$ with fluctuating hydrodynamic interactions, $\hsf > 0.25$.

Extrapolated values of $U_{\eta R}^{\theta}$ obtained from the current simulations, for each \hsH, have been averaged along with the error bars to obtain  $U_{\eta R}^{\theta} = 1.49 \pm 0.10$, which is in close agreement with the experimental value of $1.49 \pm 0.06$ reported by \citet{Miyaki1980}, and with the simulation result of $1.47 \pm 0.15$ predicted by~\citet{Torre84} using Monte Carlo rigid body simulations. A comparison between $U_{\eta R}^{\theta}$ predictions from current simulations with results of the simulations of \citet{Kroger20004767} is also shown in~\cref{fig:uetartheta}. It is clear that the scatter in the values obtained from an extrapolation of finite shear rate data is significantly more than that obtained using the method adopted in the present work.

\subsection{\label{subsec:uetar} Solvent quality crossover of $U_{\eta \! R}$}

\begin{figure*}[!htbp]
\begin{center}
\begin{tabular}{cc}
\resizebox{0.39\linewidth}{!} {\includegraphics*{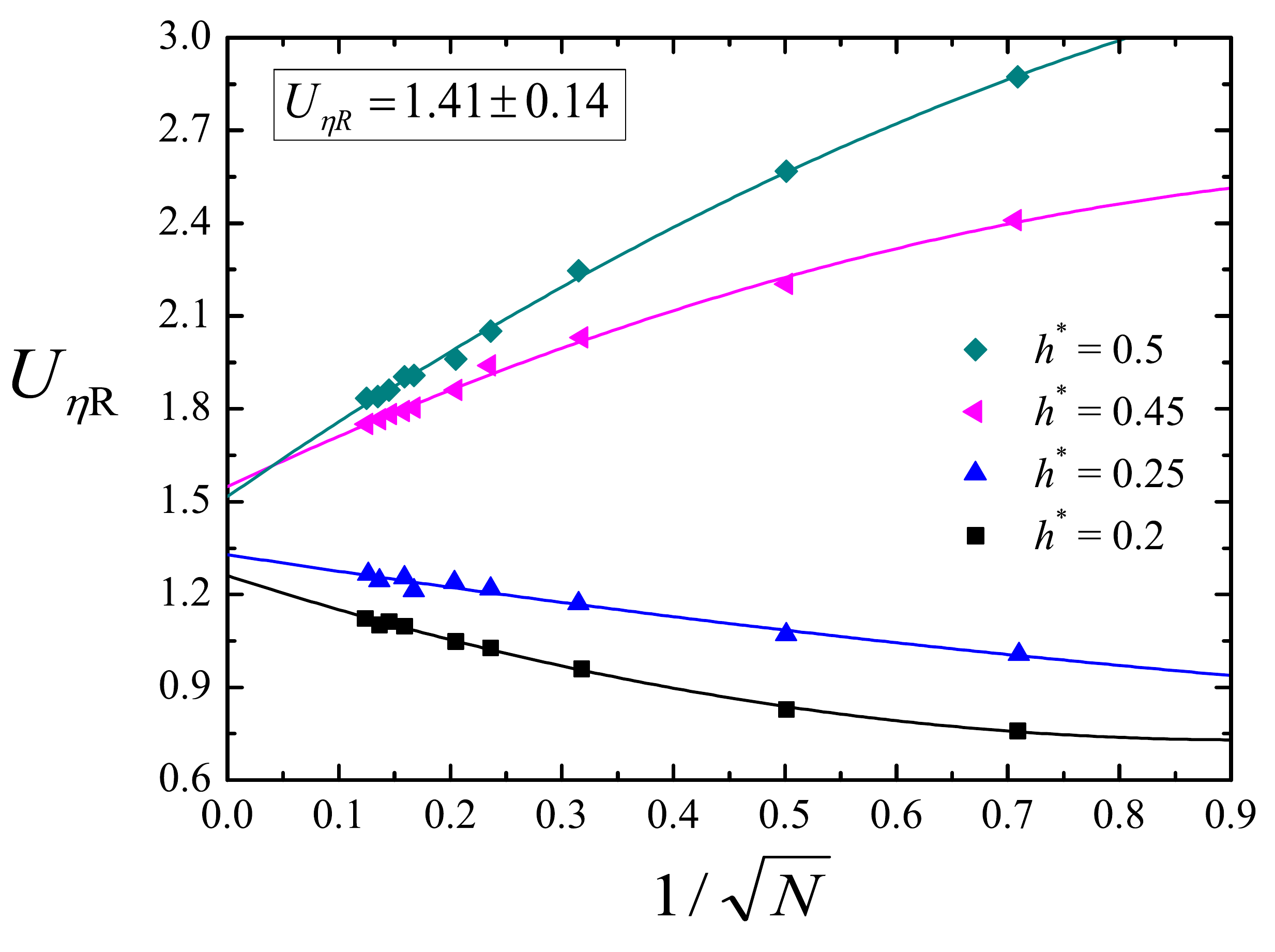}}&
\resizebox{0.39\linewidth}{!} {\includegraphics*{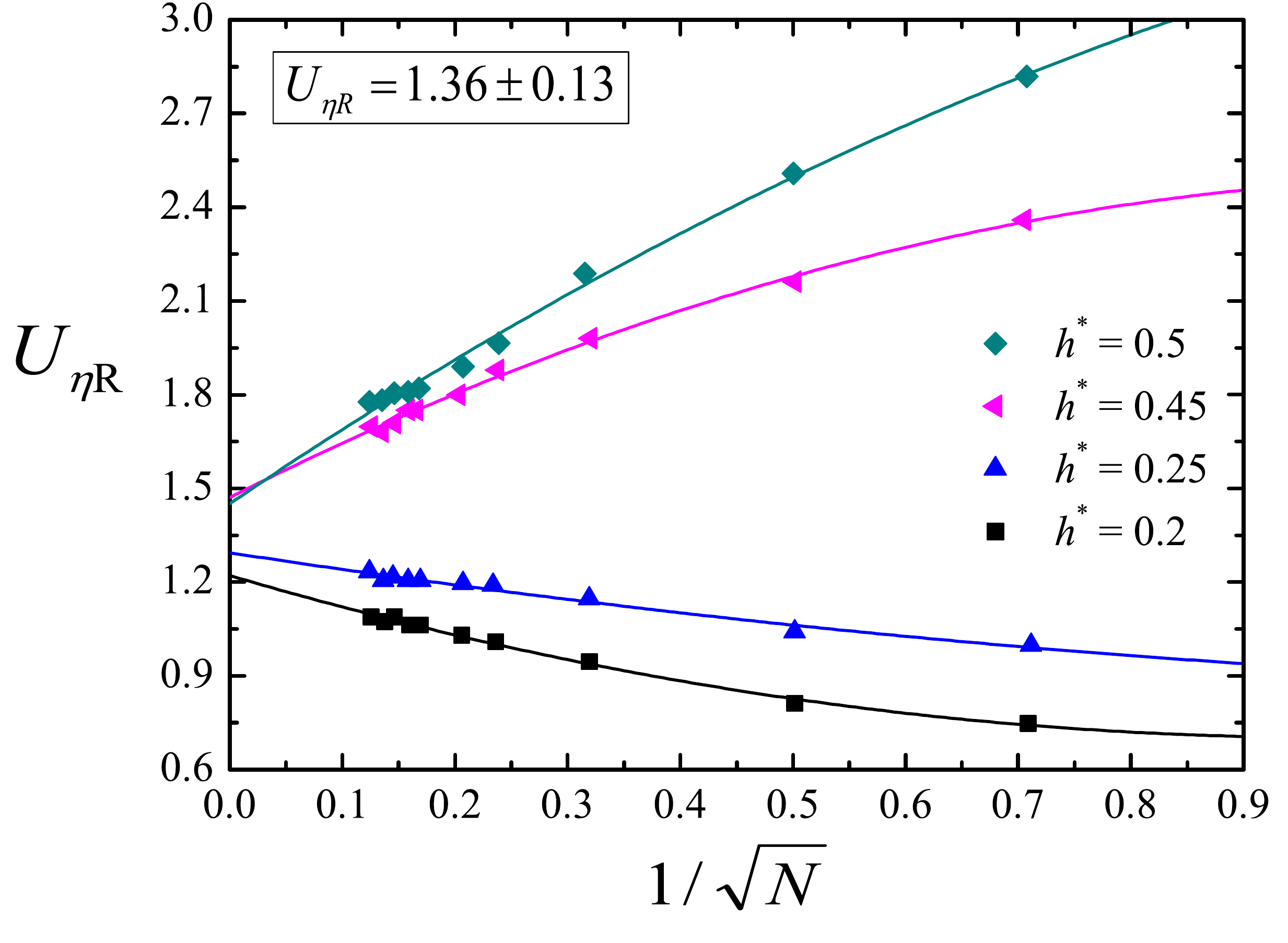}}\\
(a) & (b)  \\
\resizebox{0.39\linewidth}{!} {\includegraphics*{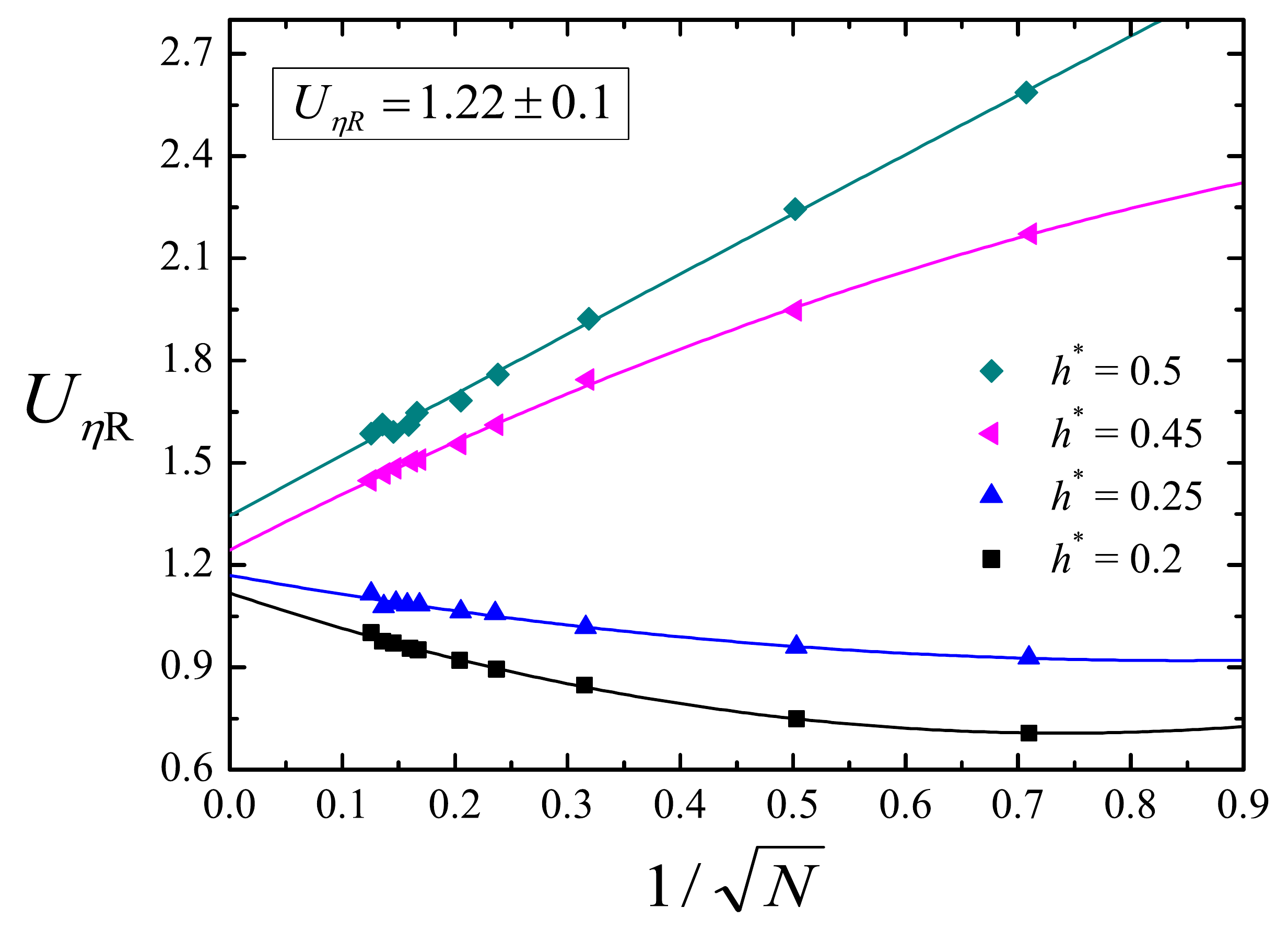}}&
\resizebox{0.39\linewidth}{!} {\includegraphics*{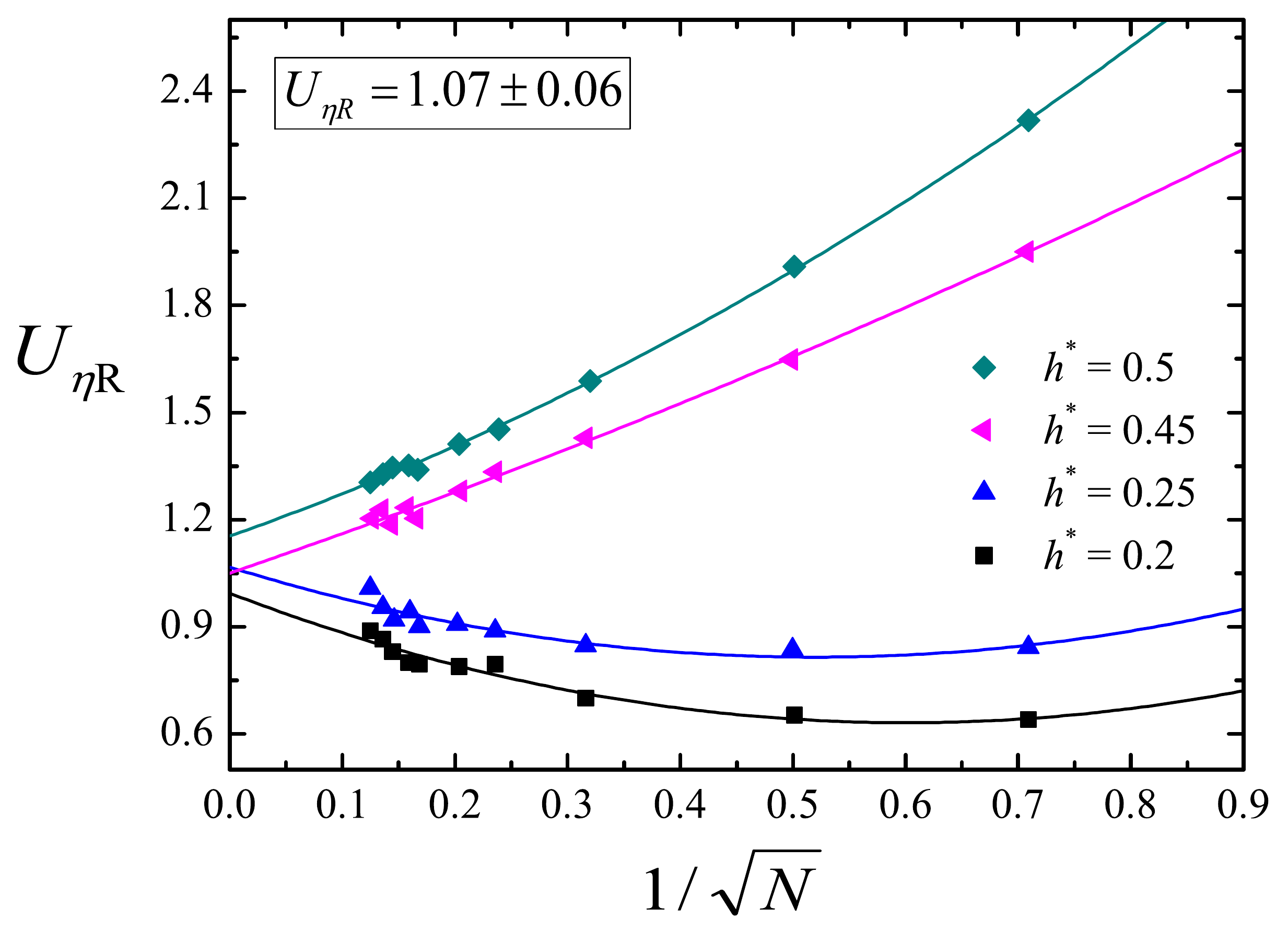}}\\
(c) & (d)
\end{tabular}
\end{center}
\vskip-10pt
\caption{\label{fig:extrapolation} Universal viscosity ratio $U_{\eta R}$  for  good solvents at fixed values of solvent quality: (a) $z$ = 0.001, (b) $z$ = 0.1, (c) $z$ = 1, and (d) $z$ = 5. The solid lines are second order polynomial fits to the BD simulations data at different values of $h^{*}$. Legends indicate extrapolated values in the long chain limit. Note that for all the simulations reported here, the parameter $K$ (related to the range of the potential, $d^{*}$) has been set equal to $1$, since the results do not depend on the value of $K$ in the limit \Ntoinf\ (Supporting Information). }
\end{figure*}

The present technique of extrapolating finite chain data to the long chain limit, while simultaneously keeping $h^{*}$ and $z$ constant, leads to asymptotic predictions of the crossover behaviour of flexible chains in the non-draining limit.
\cref{fig:extrapolation} displays the results of adopting this procedure to predict the crossover behaviour of $\Uer$. At each value of $z$, data is accumulated at fixed values of $h^{*}$ for several values of chain length $N$. The mean of the extrapolated values of $\Uer$ in the long chain limit, for the different $h^{*}$, is considered to be the universal value of $\Uer$ at that value of $z$. Legends in Figures.~(a) to~(d) of~\cref{fig:extrapolation} indicate the asymptotic values of the universal ratio obtained at the respective values of $z$.

\begin{figure}[!tbp]
\begin{center}
\resizebox{0.9\linewidth}{!} {\includegraphics*{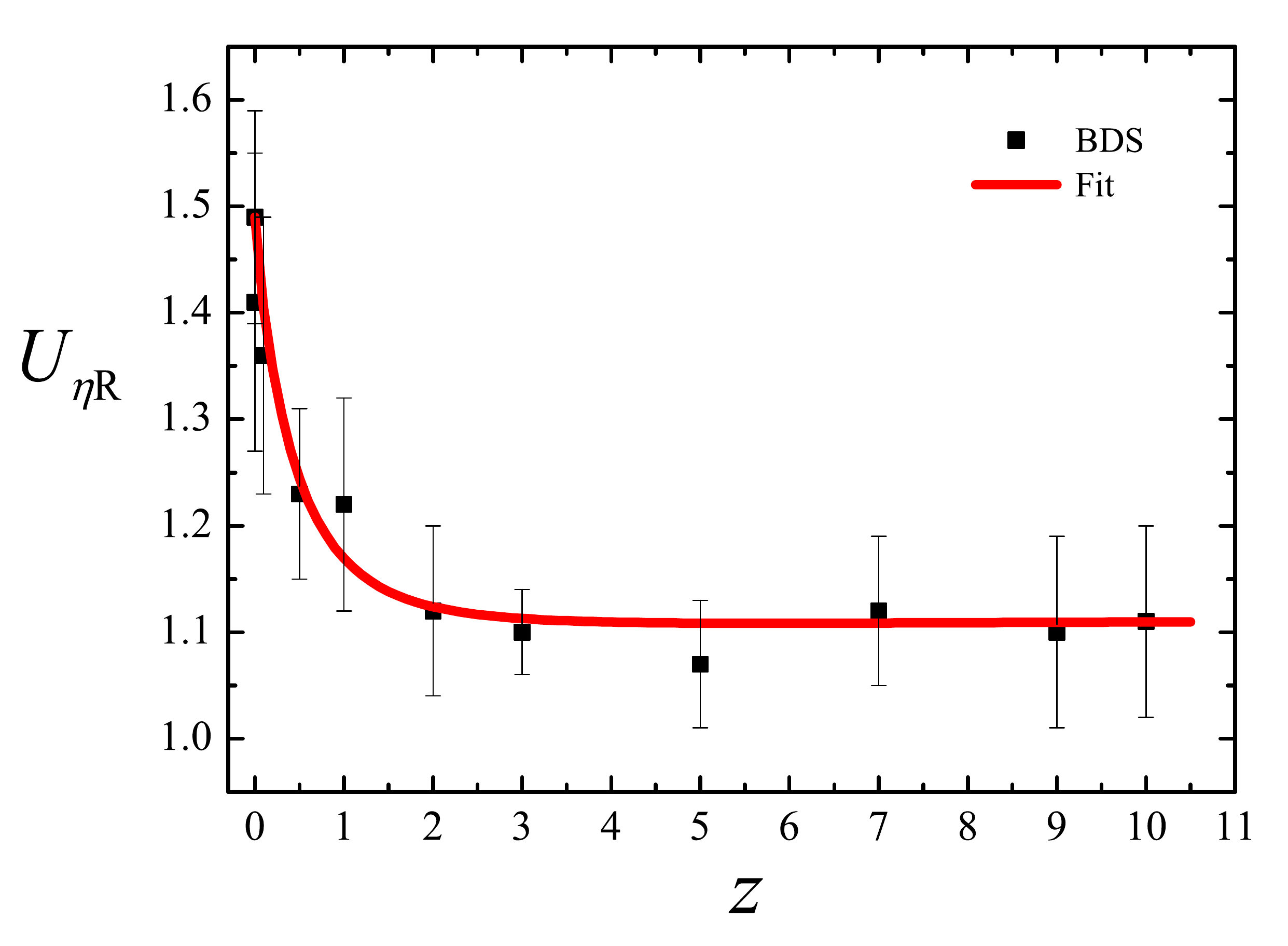}}
\end{center}
\caption{Universal viscosity ratio \Uer\ as a function of the solvent quality parameter $z$. Black squares are results of BD simulations obtained by extrapolating finite chain data to the long chain limit, as shown in~\cref{fig:extrapolation}. The solid curve is a fit to the simulation data with the expression given in~\cref{eq:uer1}.}
\label{fig:uetarvsz_sim}
\end{figure}

\cref{fig:uetarvsz_sim} displays the dependence on $z$ of the asymptotic values of  \Uer\ obtained in this manner. Starting at $U_{\eta R}^{\theta}=1.49 \pm 0.1$ at $z=0$, the universal ratio appears to decrease rapidly with increasing values of $z$, levelling off to an excluded volume limit value of $U_{\mathrm{\eta R}}^{\infty}=1.1 \pm 0.1$ for $z \gtrsim 5$.
Experimental observations of the dependence of the Flory-Fox constant on solvent quality for a number of different polymer-solvent systems have been summarised in the recent review by~\citet{Jamieson2010}. The general consensus appears to be that ${\Phi}$ decreases rapidly with increasing solvent quality, and with increasing molecular weight in good solvents. The behaviour displayed in \cref{fig:uetarvsz_sim} is in agreement with the qualitative trend expected from experimental observations~\cite{Jamieson2010}. Further, the value $U_{\mathrm{\eta R}}^{\infty}=1.1 \pm 0.1$ is in excellent agreement with the earlier prediction of $1.11 \pm 0.10$ by~\citet{Torre91} in the good solvent limit.

As will be discussed in greater detail in~\cref{subsec:alphaetavsz} below, the dependence of the swelling \aeta\ on the solvent quality $z$, predicted by Brownian dynamics simulations, can be represented by a functional form identical to that for \ag\ in~\cref{eq:ag}, with values of the parameters $a$, $b$ and $c$ as given in~\cref{tab:fitpars}. The value of the exponent $m$, however, is the same in the expressions for both the crossover functions \aeta\ and \ag, since (as can be seen from~\cref{eq:uetaratio1}), this must be true in order for $\Uer$ to level off to a constant value for large values of $z$,  as observed in the BD simulations displayed in~\cref{fig:uetarvsz_sim}.
Using the functional forms for \aeta\ and \ag, and~\cref{eq:uetaratio1}, it follows that,
\begin{equation}
\label{eq:uer1}
{\Uer} =  {U_{\eta R}^{\theta}} \left(  \frac{1 + a_{\eta} z + b_{\eta} z^{2} + c_{\eta} z^{3}}{1 + a_{g} z + b_{g} z^{2} + c_{g} z^{3}}\right)^{3m/2}
\end{equation}
where, the suffixes on the parameters $a$, $b$ and $c$ indicate the relevant crossover function. The red curve in~\cref{fig:uetarvsz_sim} is a fit to the BD simulation data using~\cref{eq:uer1}, along with ${U_{\eta R}^{\theta}} = 1.49$, and the appropriate values for the fitting parameters listed in~\cref{tab:fitpars}. Clearly the fit is very good, as can be expected from the excellence of the fits for the crossover functions for  \aeta\ and \ag.

\begin{figure}[!tbp]
\begin{center}
\resizebox{0.9\linewidth}{!} {\includegraphics*{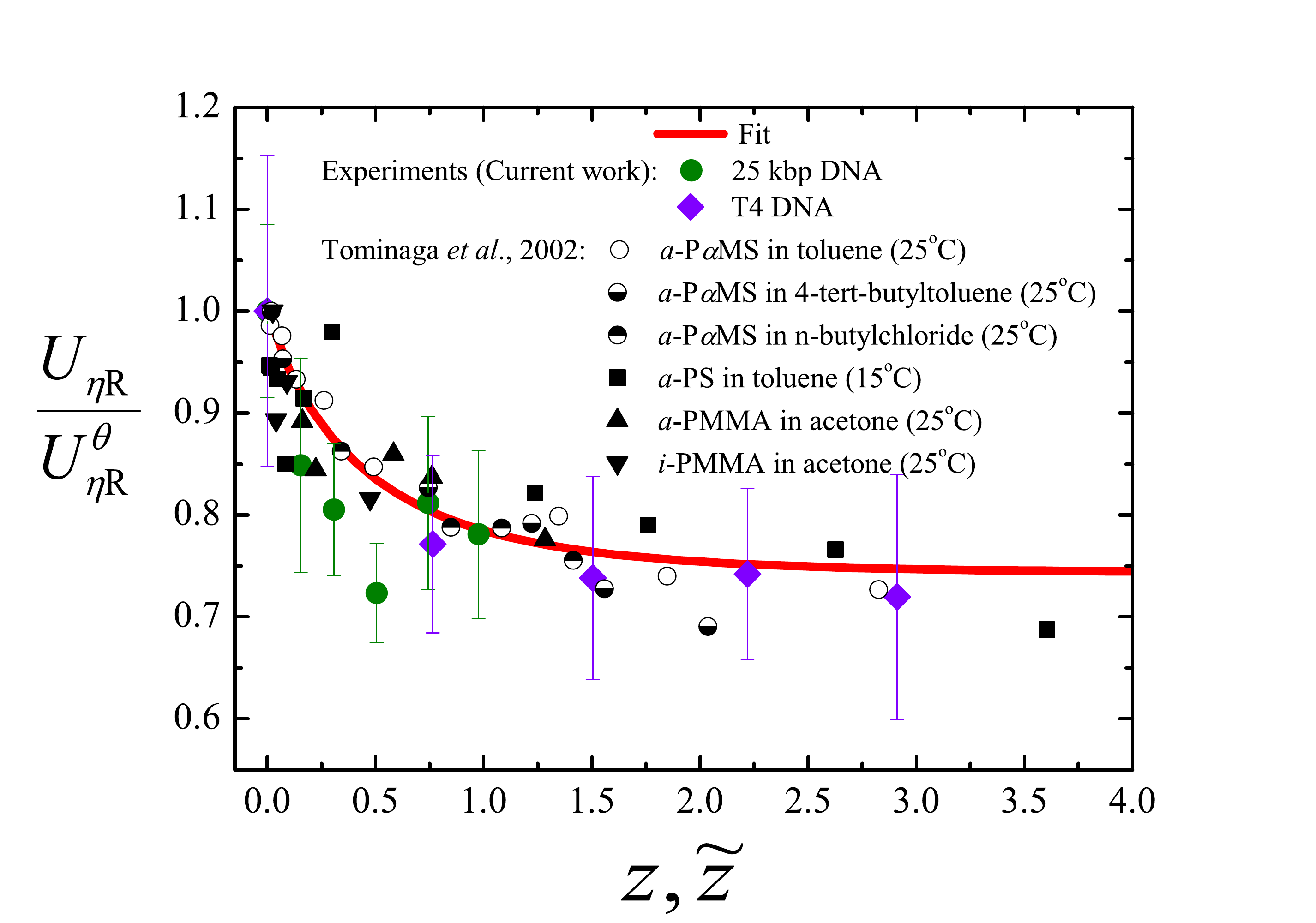}}
\end{center}
\caption{Comparison of the experimentally determined dependence of $({\Uer}/{U_{\eta R}^{\theta}})$ on solvent quality with the prediction of Brownian dynamics simulations. The solid curve is a fit to BD simulation data using \cref{eq:uer1}.}
\label{fig:uetarvsz_exp}
\end{figure}

\citet{Tominaga20021381} have reported experimental measurements of the dependence of \aeta\ on $\tilde z$,  and have also plotted $\log \aeta^{3}$ versus $\log \ag^{3}$, for a number of different wormlike polymer-solvent systems. Consequently, using~\cref{eq:uetaratio1}, the dependence of $({\Uer}/{U_{\eta R}^{\theta}})$ on $\tilde z$ can be determined for all the experimental systems studied in Ref.~\citenum{Tominaga20021381}. As discussed earlier in~\cref{subsec:ivisc}, this ratio can also be determined, as a function of $\tilde z$,  for the 25 kbp and T4 DNA samples studied here. \cref{fig:uetarvsz_exp} displays the data extracted from  \citet{Tominaga20021381} in this manner, alongside the DNA measurements from the current work, and the curve fit to the BD simulations data for $({\Uer}/{U_{\eta R}^{\theta}})$ as a function of $z$. The experimental data can be seen to be scattered around the BD simulation curve, and closely follow the trend of rapid decrease in $({\Uer}/{U_{\eta R}^{\theta}})$ with increasing solvent quality. In particular, experimental measurements for the two DNA samples lie close to the observations for synthetic polymer-solvent systems, and to the BD simulation curve. This suggests that the expectation of quasi-two-parameter theory, that the functional dependence of $({\Uer}/{U_{\eta R}^{\theta}})$ on $\tilde z$ to be identical to that of its dependence on $z$, is justifiable.

\begin{table}[!tbp]
\addtolength{\tabcolsep}{2pt}
\caption{\label{tab:fitpars} Values of the parameters $a$, $b$, $c$ and $m$ in the functional form $f(z) = (1 + az + bz^{2} + cz^{3})^{m/2}$ used to fit the Brownian dynamics simulations data for the crossover functions \ag, \aeta\ and \ah.}
\vspace{2pt}
\centering
\begin{tabular}{cccc}
\hline
    & \ag\ & \aeta\ & \ah\  \\
\hline
$a$ & 9.5286      & 5.4475 $\pm$ 1.776      &  9.528                      \\
$b$ & 19.48 $\pm$ 1.28           & 3.156 $\pm$ 1.982      &  19.48                     \\
$c$ & 14.92 $\pm$ 0.93           & 3.536 $\pm$ 0.277    &  14.92                    \\
$m$ & 0.133913 $\pm$ 0.0006      & 0.1339                  &  0.0995 $\pm$ 0.0014  \\
\hline
\end{tabular}
\end{table}

For large values of $z$, \cref{eq:uer1} implies that the excluded volume limit value of the ratio, from fitting Brownian dynamics simulations is, $({\Uerinf}/{U_{\eta R}^{\theta}}) =  \left(  {c_{\eta}}/{c_{g}}\right)^{3m/2} = 0.749$. Experimental measurements appear to indicate a value of the ratio, $\Phi/\Phi_{0} \approx 0.773$~\cite{Jamieson2010}, while the Monte Carlo rigid body simulations of~\citet{Torre91} lead to $\Phi/\Phi_{0} \approx 0.76$.

\subsection{\label{subsec:alphaetavsz} Swelling of the viscosity radius}

\begin{figure}[tbp]
\begin{center}
\resizebox{0.9\linewidth}{!} {\includegraphics*{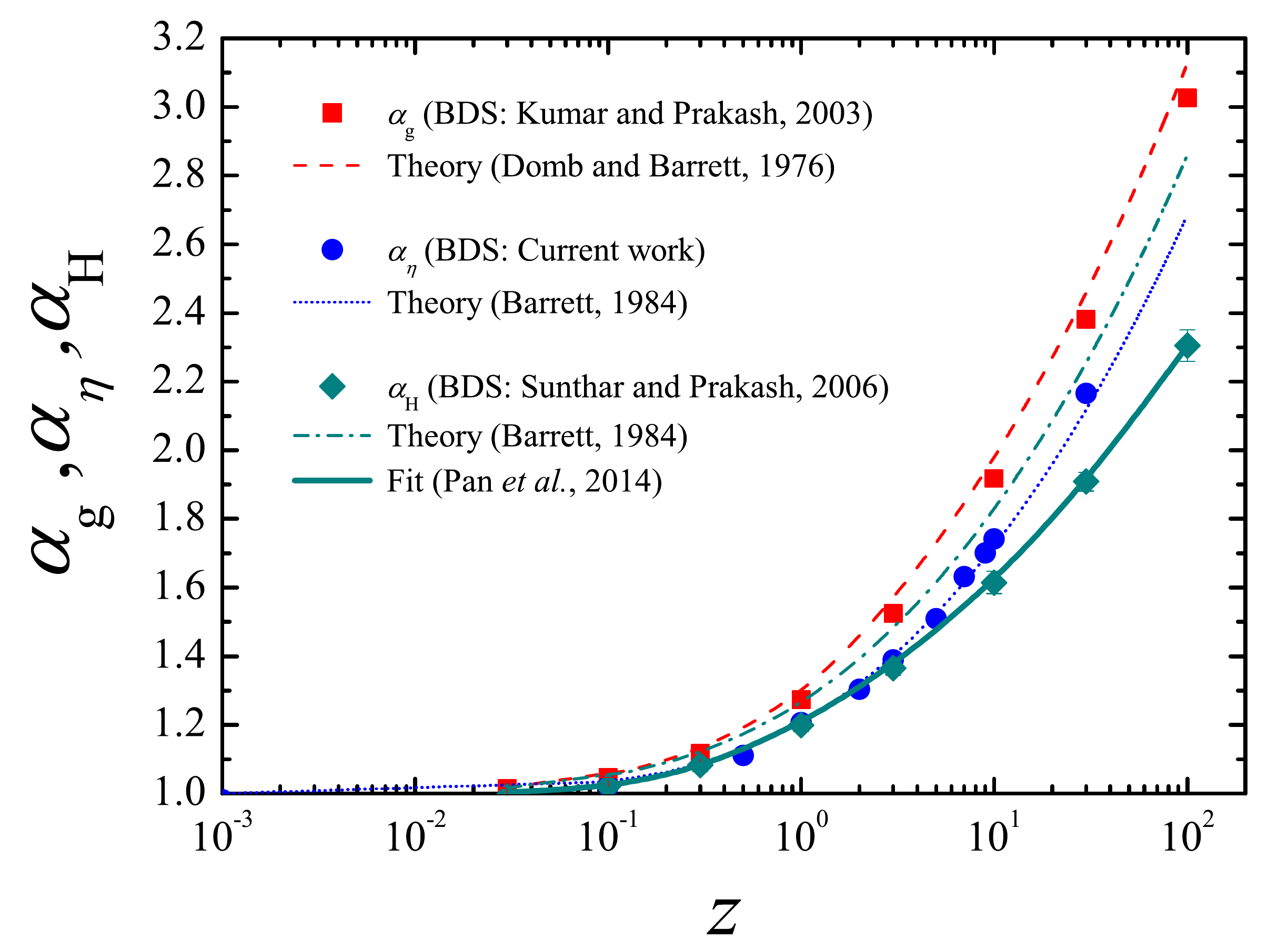}}
\end{center}
\vskip-20pt
\caption{Universal crossover scaling functions for  \ag, \ah, and \aeta\ predicted by BD simulations. Filled blue circles are the predictions of \aeta\ in the current work, while filled red squares and the filled green diamonds are previous BD simulation predictions of \ag~\cite{Kumar20037842} and  \ah~\cite{SunRav06-epl}, respectively. The solid green line is an analytical fit to simulation data for \ah\ with the functional form $f(z) =  \left(1+ az + bz^{2} + cz^{3} \right)^{m}$, where the constants $a$, $b$, $c$ and $m$, are as given in~\cref{tab:fitpars}. Predictions by the Domb-Barrett equation~\cite{dombbarrett76} for \ag\ (red dashed curve), and the Barrett equations~\cite{Barrett1984} for \ah\ (green dot-dashed curve) and \aeta\ (blue dotted curve) are also displayed. }
\label{fig:compare}
\end{figure}

The prediction of the swelling $\aeta$ as a function of $z$ from current simulations, using~\cref{eq:aeta1}, is displayed in~\cref{fig:compare} by the filled blue symbols. For comparison, previous BD predictions by our group of \ag\ (red symbols) and \ah\ (green symbols), and the crossover functions predicted by the Domb-Barrett and Barrett theories have also been displayed in~\cref{fig:compare}. The solid green line is a fit to the BD simulation data for \ah\ using the functional form $f(z) = \left( 1+ az + bz^{2} + cz^{3} \right)^{m}$, with the parameters $a$, $b$, $c$ and $m$ listed in~\cref{tab:fitpars} (as reported previously in Ref.~ \citenum{Pan2014339}). As mentioned earlier  in~\cref{subsec:uetar}, we have used this functional form to fit the data for \aeta\ as well, with the constraint that $m_{\eta} = m_{g}$.

The difference between the static scaling function \ag\ and the dynamic scaling functions \ah\ and \aeta\ is clearly visible, with the dynamic scaling function for \ah, in particular, showing a slow approach to the asymptotic scaling exponent at large $z$. The agreement of the Barrett equation for \aeta, based on pre-averaged hydrodynamic interactions, with BD simulations that account exactly for fluctuating hydrodynamic interactions, implies that the influence of fluctuations on \aeta\ are not significant, as noted by~\citet{Yamakawa1995}. On the other hand, the disagreement of the Barrett equation for \ah, with exact BD simulations, is due to the more pronounced influence of fluctuating hydrodynamic interactions on \ah. As mentioned previously, the Barrett equation for \ah\ is unable to predict experimental observations, while the BD simulations are quantitatively accurate~\cite{SunRav06-epl}. Interestingly, the curves for \ah\ and \aeta\ coincide for values of $z \lesssim 5$. This is the reason that the Barrett equation for \aeta\ is often used to describe experimental data for \ah. However, the curves depart from each other for larger values of $z$, with the curve for \aeta\ becoming parallel to that for \ag. This is to be expected since experimental observations suggest that $\Uer$ is a universal constant in $\theta$-solutions and in the excluded volume limit, and as a result, \cref{eq:uetaratio1} implies that $\aeta$ must scale linearly with $\ag$ for large $z$.

\begin{figure}[tbp]
\begin{center}
\resizebox{1.1\linewidth}{!} {\includegraphics*{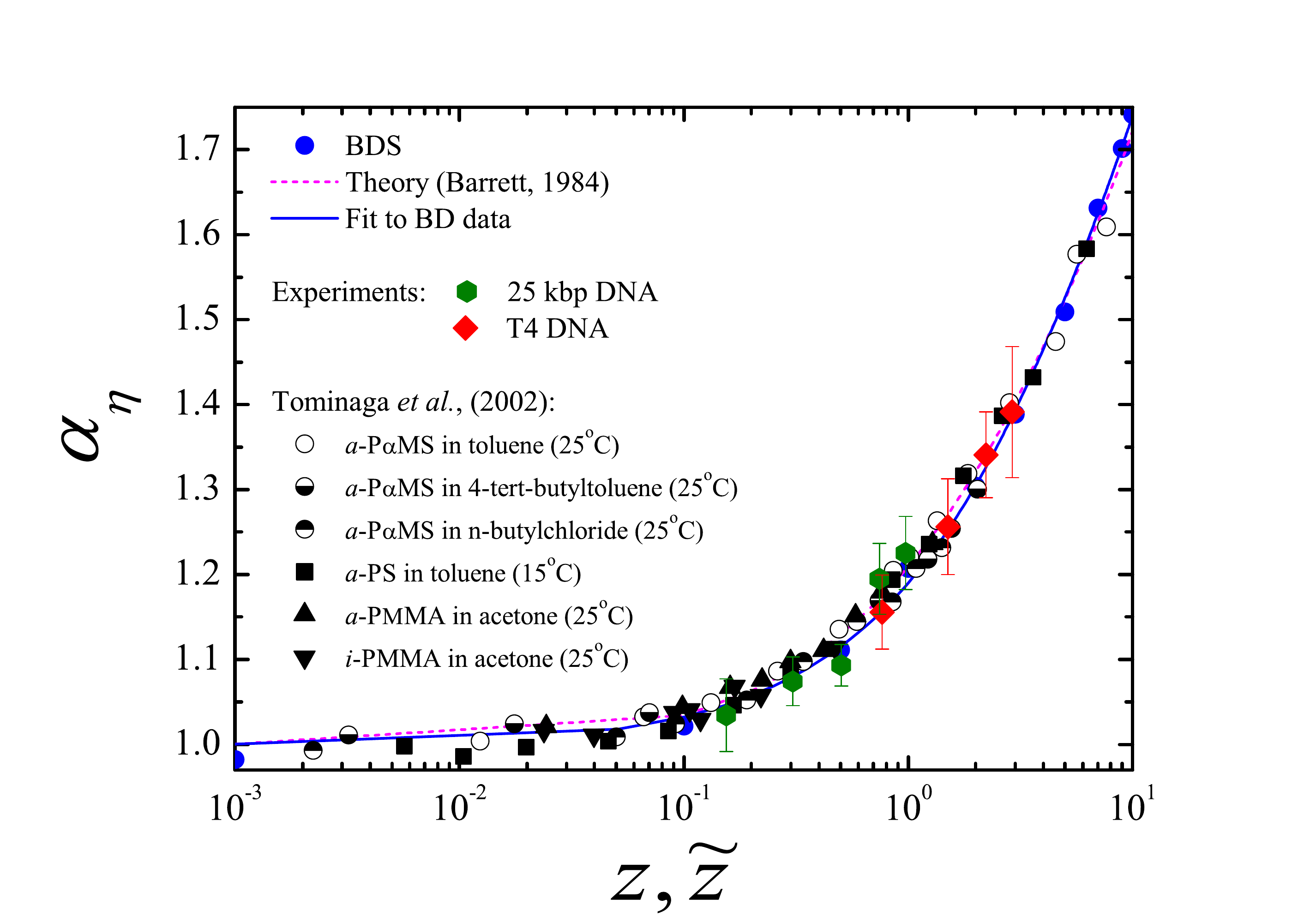}}
\end{center}
\vskip-20pt
\caption{Crossover swelling of the viscosity radius from $\theta$ to good solvents. Experimental measurements of the swelling of 25 kbp and T4 DNA  are represented by the filled hexagons and diamonds, respectively, while the remaining symbols represent data on various synthetic wormlike polymer-solvent systems collated in~\citet{Tominaga20021381} The filled blue circles are the predictions of the current BD simulations. The solid line represents a fit to the BD data with the functional form $f(z) =  \left(1+ az + bz^{2} + cz^{3} \right)^{m}$, where the constants $a$, $b$, $c$ and $m$, are as given in~\cref{tab:fitpars}, while the dotted red line is the prediction of the  Barrett equation~\cite{Barrett1984} for \aeta.}
\label{fig:alphaeta}
\end{figure}

Experimental measurements of \aeta\ as a function of the scaled excluded volume parameter $\tilde z$, obtained in the present work for 25 kbp and T4 DNA, are plotted alongside the predicted dependence of \aeta\ on $z$ by current BD simulations, in \cref{fig:alphaeta}.  Previous measurements of \aeta\ as a function  $\tilde z$, reported in~\citet{Tominaga20021381}  for solutions of synthetic wormlike polymers, are also displayed in \cref{fig:alphaeta} for the purpose of comparison. Here again, the assumption of quasi-two-parameter theory that \aeta\ depends identically on $z$ and $\tilde z$ is seen to be validated. The excellent agreement between the swelling of DNA, and synthetic polymer-solvent systems implies that the swelling of the viscosity radius of DNA, in dilute solutions with excess salt, is universal.

\begin{figure}[tbp]
\begin{center}
\resizebox{1.1\linewidth}{!} {\includegraphics*{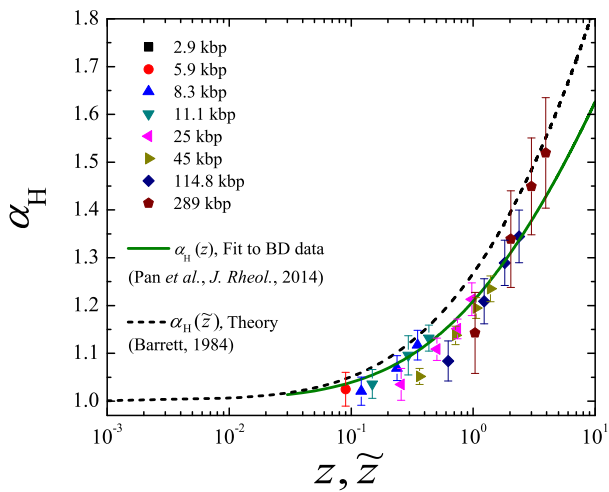}}
\end{center}
\vskip-20pt
\caption{Crossover swelling of the hydrodynamic radius from $\theta$ to good solvents. Symbols represent experimental measurements of the swelling of DNA, of various molecular weights, as a function of the scaled excluded volume parameter $\tilde z$. The solid line is a fit to previous BD simulations data~\cite{SunRav06-epl} with the functional form $f(z) =  \left(1+ az + bz^{2} + cz^{3} \right)^{m}$, where the constants $a$, $b$, $c$ and $m$, are as given in~\cref{tab:fitpars}, while the dashed line is the prediction of the  Barrett equation~\cite{Barrett1984} for \ah.}
\label{fig:alphah}
\end{figure}

As mentioned previously, \citet{Pan2014339} have used dynamic light scattering to determine the dependence of the swelling ratio \ah\ on $z$, assuming that DNA is a flexible molecule at the molecular weights that were considered. In~\cref{fig:alphah}, the data for \ah\ (from Ref.~\citenum{Pan2014339}) is replotted as a function of $\tilde z$, by taking into account the wormlike character of DNA (see Supporting Information for details). The collapse of the data for DNA onto master plots, for both \aeta\ and \ah\ in~\cref{fig:alphaeta} and~\cref{fig:alphah}, respectively, validates the estimation by \citet{Pan2014339} of the $\theta$-temperature for DNA solutions in the presence of excess salt to be $T_{\theta} \approx 15 \degC$, and the procedure given in the Supporting Information for the determination of the solvent quality $\tilde z$, at any given molecular weight $M$ and temperature $T$. Further, the agreement between experimental observations and BD simulations suggests that the simulation framework used here is highly suited to obtain accurate predictions of universal behaviour of dilute polymer solutions in the entire solvent quality crossover regime.

There has been some discussion in the literature recently, based on Monte Carlo simulations, regarding the use of double-stranded DNA as a model polymer to capture long chain universal behaviour, due to the structural rigidity of the double helix~\cite{Tree2013}. The results displayed in~\cref{fig:uetarvsz_exp}, \cref{fig:alphaeta} and~\cref{fig:alphah} indicate that double-stranded DNA is indeed a model polymer, over a wide range of molecular weights.

\section{\label{sec:conc} Conclusions}

The intrinsic viscosities of dilute DNA solutions, of two different molecular weight samples (25 kbp and T4 DNA), have been measured at different temperatures in a commonly used solvent under excess salt conditions (Tris-EDTA buffer with 0.5 M NaCl). The measurements have been used to calculate the swelling of the viscosity radius $\alpha_{\eta}$ and the universal viscosity ratio $\Uer$, as a function of the solvent quality $\tilde z$. In parallel, universal predictions of these crossover functions have been obtained with the help of BD simulations that incorporate fluctuating hydrodynamic interactions, in the non-draining limit.

The experimental measurements of \Uer\  and \aeta\ for the DNA solutions are found to collapse onto  previously reported data for synthetic polymer-solvent systems, and onto the current BD simulations predictions. The close agreement between prior experiments, current experiments and simulations suggests that: (i) DNA solutions in the presence of excess salt exhibit universal behaviour in line with similar observations for synthetic polymer solutions, and (ii) the model used here incorporates all the important mesoscopic physics necessary to capture the universal behaviour of equilibrium static and dynamic properties of dilute polymer solutions. In particular, the model enables the elucidation of the role played by hydrodynamic interactions in determining the differences in the observed scaling of static and dynamic crossover functions.

\section*{Acknowledgements}

This research was supported under Australian Research Council's Discovery Projects funding scheme (project number DP120101322). We are grateful to Douglas E. Smith and his group in the University of California, San Diego, for preparing the special DNA fragments and to Brad Olsen, MIT, for the stab cultures containing them. The authors would like to thank M. K. Danquah (formerly at Monash University) for providing laboratory space for storing DNA samples, and for the instruments and facilities for extracting DNA. We also acknowledge funding received from the IITB-Monash Research Academy. We thank the anonymous referees for helpful suggestions that have improved the quality of the paper.\\

\section*{Supporting Information}

Supporting information contains table of properties of DNA molecules; solvent details and estimation of DNA concentration; plots for determination of zero shear rate viscosity from measurements of viscosity at different finite shear rates; table of zero shear rate viscosity values for various concentrations and temperatures; determination of the chemistry dependent constant $k$, and mapping between $T$ and $M$, and $\tilde z$; stochastic differential equation for bead positions; precise forms of the excluded volume potential and hydrodynamic interaction tensor; features of the Brownian dynamics integration algorithm; Fixman's expressions for $\bar H_{\mu\nu}$ and $\langle {\hat \CS (t)} \rangle_{\mathrm{eq}}$; integration of the correlation functions.

\bibliography{alphaeta}

\clearpage
\section*{Supporting Information}

\renewcommand{\thesection}{S\arabic{section}}
\setcounter{section}{0}

\section{DNA samples}

\begingroup
\small
\begin{table*}[t]
  \caption{\label{tab:relaxationtime} Representative properties of the 25 kbp and T4 DNA
    used in this work (reproduced from Table I of Ref.~\citenum{Pan2014339}). $L$  is the contour length, $N_{\mathrm{k}}$ is the number of Kuhn steps, and $\Rgth$ is the radius of gyration at
    the $\theta$ temperature. The two relaxation times at the $\theta$
    temperature are defined by $\lambda_{\mathrm{D}}^{\theta} =
    {\left(\Rgth\right)}^{2} / \Dth$, where \Dth\ is the measured
    diffusion coefficient under $\theta$ conditions,  and
    $\lambda_{\eta}^{\theta} = (M \eta_{\mathrm{p0}}) / (c
    N_{\mathrm{A}} k_{\mathrm{B}} T)$, where $c$ is the concentration, $N_{\mathrm{A}}$ the Avagadro number, and $k_{\mathrm{B}}$ the Boltzmann constant. While
    $\lambda_{\mathrm{D}}^{\theta}$ is evaluated at $c/c^{*} = 0.1$,
    $\lambda_{\eta}^{\theta}$ is calculated at $c/c^{*} = 1$.}
\vskip10pt
\begin{tabular}{ c  c  c  c  c  c  c }
\hline
DNA Size (kbp)      & $M$ ($\times 10^{6}$ g/mol)
            & $L (\mu)$
            & $N_{\mathrm{k}}$
            & $R_{\mathrm{g}}^{\theta}$ (nm)
            & $\lambda_{\mathrm{D}}^{\theta}$ ($\times$10$^{-3}$ s)
            & $\lambda_{\eta}^{\theta}$ ($\times$10$^{-1}$ s)
\\
\hline
\hline
25
            & 16.6
            & 9
            & 85
            & 376
            & 197
            & 1.19
\\
\hline
165.6
            & 110
            & 56
            & 563
            & 969
            & --
            & 51.9
\\
\hline
\end{tabular}
\vskip10pt
\end{table*}
\endgroup

Typical properties of the DNA molecules used in this work, such as the molecular weight, contour length, number of Kuhn steps, etc., are tabulated in Table I, and have been reproduced here from a similar Table in \citet{Pan2014339} The T4 and 25 kbp DNA samples were dissolved in a solvent containing 10 mM Tris (\#T1503, Sigma-Aldrich), 1 mM EDTA (\#E6758, Sigma-Aldrich) and 0.5 M NaCl (\#S5150, Sigma-Aldrich), which was also used for preparing subsequent dilutions. The solvent has a viscosity of 1.01 mPa.s at 20\degC, which is approximately equal to the  viscosity of water.

For T4 linear genomic DNA, with an anticipated purity of high order, the concentration of 0.24 mg/ml specified by the company was used. For the 25 kbp linear DNA, the concentration of DNA (0.441 mg/ml) was determined by both UV-VIS spectrophotometry (\#UV-2450, Shimadzu) and agarose gel electrophoresis, the latter by comparing with a standard DNA marker (\#N0468L, New England Biolabs). The $A_{260}/A_{280}$ and $A_{260}/A_{230}$ ratios were 1.92 and 2.1 respectively, the latter indicating absence of organic reagents like phenol, chloroform etc.~\cite{Sambrook2001}, and suggesting an overall good quality of the DNA sample, as noted earlier by \citet{Laib20064115}

\begin{figure}[tbp]
\begin{center}
\begin{tabular}{c}
\resizebox{0.9\linewidth}{!} {\includegraphics*{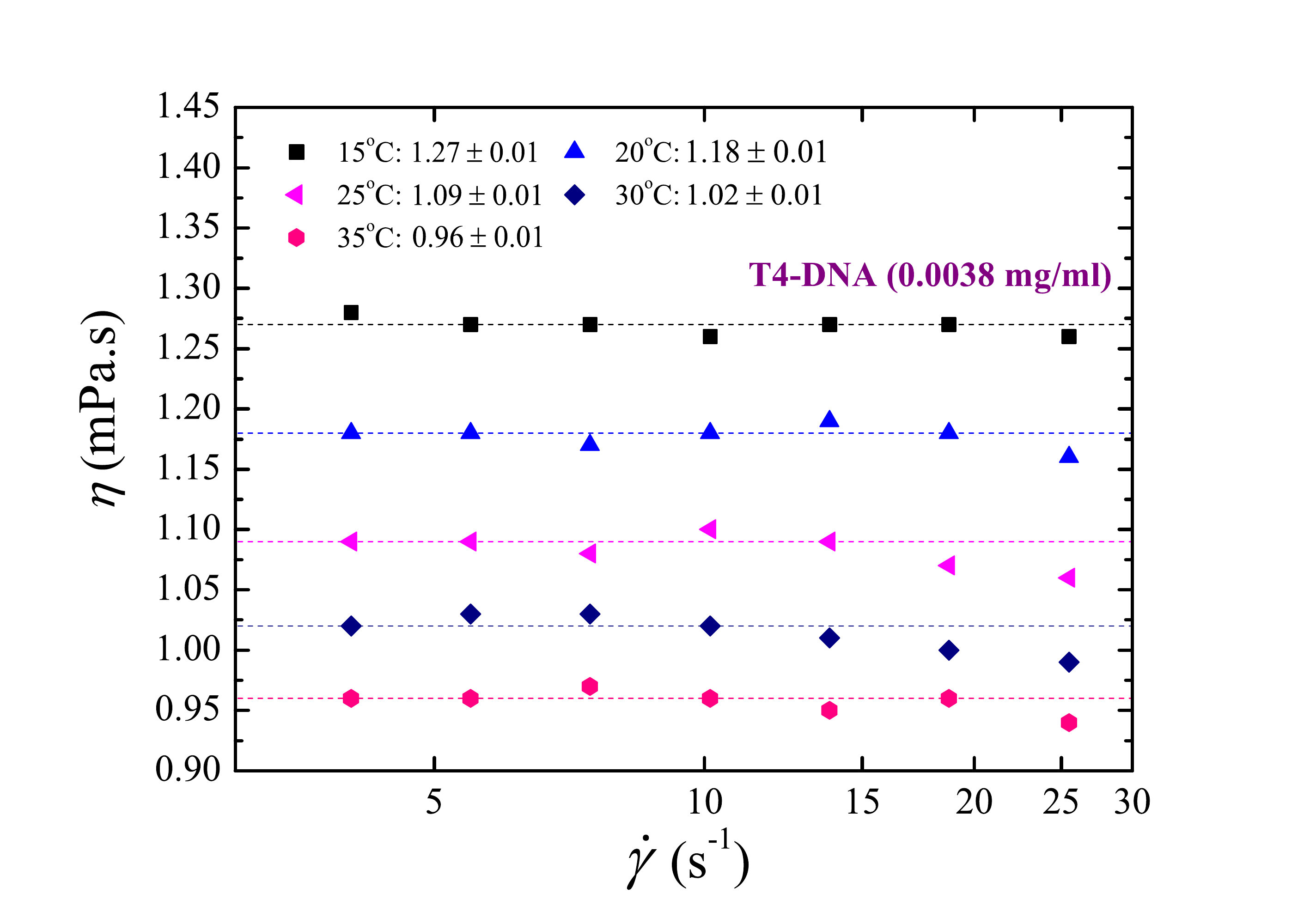}} \\
(a)  \\
\resizebox{0.9\linewidth}{!} {\includegraphics*{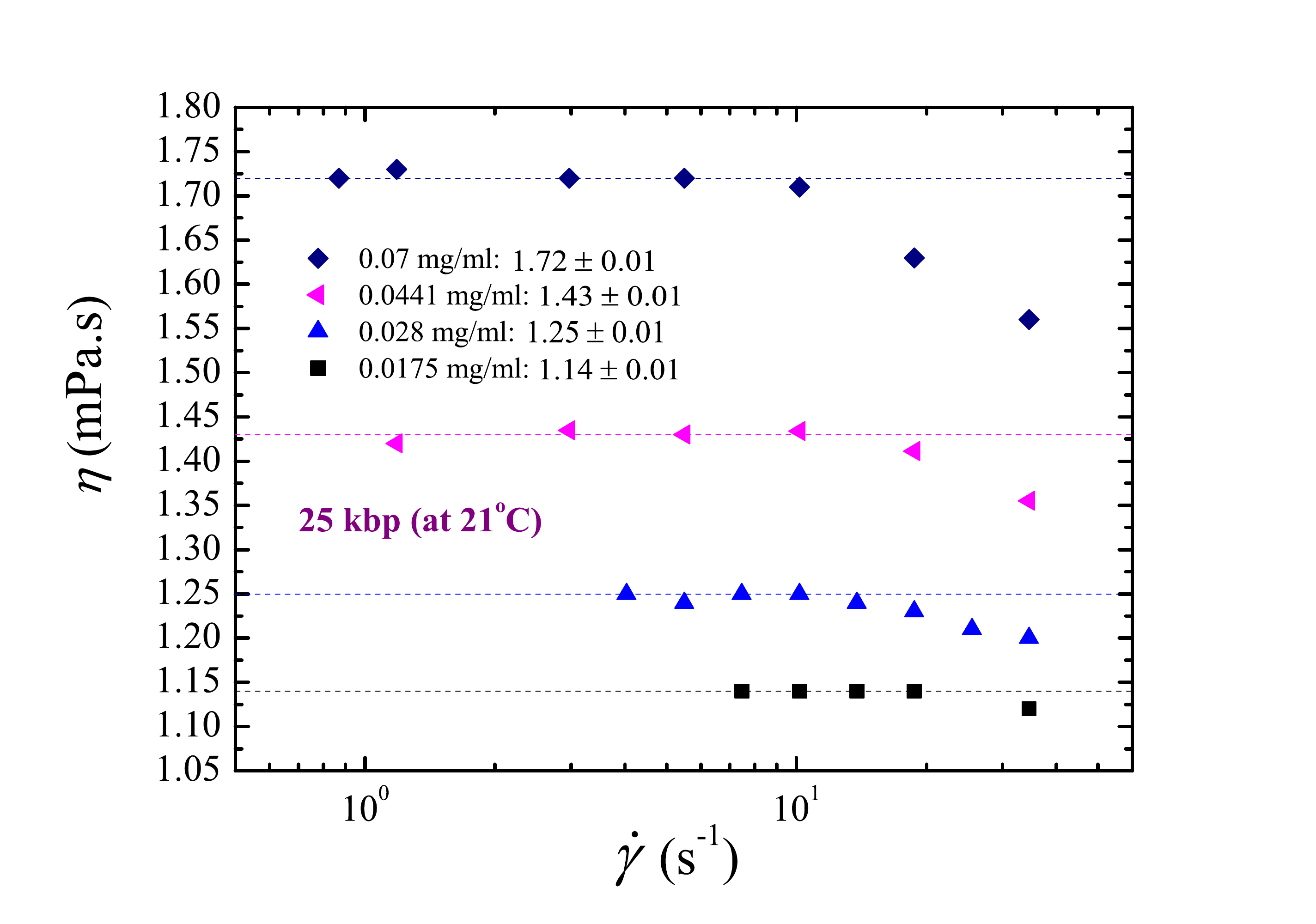}} \\
(b)  \\
\end{tabular}
\end{center}
\caption{\label{fig:rawvisc} Determination of the zero shear rate solution viscosity $\eta_{0}$. The shear rate dependence of solution viscosity $\eta$ in the region of low shear rate is extrapolated to zero shear rate (a) for T4 DNA at a fixed concentration, for a range of temperatures and (b) for 25 kbp DNA at a fixed temperature, for a range of concentrations. The extrapolated values in the limit of zero shear rate are indicated in the legends.}
\end{figure}

\section{Shear rheometry}

A Contraves Low Shear 30 rheometer with Couette geometry (1T/1T--Cup and bob; shear rate ($\dot{\gamma}$) range: 0.01--100 $s^{-1}$; temperature sensitivity: $\pm$ 0.1\degC) has been used for all the shear viscosity measurements. Recently, \citet{Heo20051117} have given a detailed description of the measuring principles underlying the Contraves rheometer. Prior to measuring the viscosity of DNA solutions, the rheometer was calibrated with Newtonian Standards (silicone oils) of known viscosities, and the zero error adjustment was carried out as described earlier in \citet{Pan2014339}

A continuous shear ramp was avoided, and to avert the problem of aggregation of long DNA chains, T4 DNA (at its maximum concentration) was kept at 65\degC\ for 10 minutes and instantly put into ice for 10 minutes~\cite{Heo20051117}. A manual delay of 30 seconds was applied at each shear rate to allow the DNA chains to relax to their equilibrium state and the sample was equilibrated for 30 minutes at each temperature. Some typically observed relaxation times are given in Table I.

\begin{table}[tbp]
\addtolength{\tabcolsep}{-2pt}
\caption{Steady state zero shear rate viscosities, $\etao$ (mPa.s) for 25 kbp and T4 DNA at various concentrations, $c$ (mg/ml) and temperatures, $T$ (\degC) in the dilute regime. Note that $T_{\theta} \approx 15 \degC$.}
\label{tab:visc}
\begin{minipage}[t]{.45\linewidth}
\vspace{0pt}
\centering
\begin{tabular}{ l  l c  c }
\multicolumn{4}{c}{25 kbp}\\
\hline
$c$ & T & $c/c^{*}$ & $\eta_{0}$ \\ [0.5ex]
\hline
\hline
0.112 & 15 & 0.91 & 2.95 $\pm$ 0.01 \\
\hline
0.07  & 15 & 0.57 & 1.76 $\pm$ 0.01 \\
      & 18 & 0.74 & 1.75 $\pm$ 0.01 \\
      & 21 & 0.85 & 1.72 $\pm$ 0.01 \\
      & 25 & 0.97 & 1.58 $\pm$ 0.02 \\
\hline
0.0441 & 15 & 0.36 & 1.53 $\pm$ 0.01 \\
       & 18 & 0.46 & 1.49 $\pm$ 0.01 \\
       & 21 & 0.54 & 1.43 $\pm$ 0.01 \\
       & 25 & 0.61 & 1.31 $\pm$ 0.01 \\
       & 30 & 0.7  & 1.27 $\pm$ 0.01 \\
       & 35 & 0.76 & 1.2 $\pm$ 0.01 \\
\hline
0.028  & 15 & 0.23 & 1.38 $\pm$ 0.01 \\
       & 18 & 0.29 & 1.31 $\pm$ 0.01 \\
       & 21 & 0.34 & 1.25 $\pm$ 0.01 \\
       & 25 & 0.39 & 1.15 $\pm$ 0.01 \\
       & 30 & 0.44 & 1.09 $\pm$ 0.01 \\
       & 35 & 0.48 & 1.01 $\pm$ 0.01 \\
\hline
0.0175  & 15 & 0.14 & 1.29 $\pm$ 0.01 \\
        & 18 & 0.18 & 1.2 $\pm$ 0.01 \\
        & 21 & 0.21 & 1.14 $\pm$ 0.01 \\
        & 25 & 0.24 & 1.05 $\pm$ 0.01 \\
        & 30 & 0.28 & 0.98 $\pm$ 0.01 \\
        & 35 & 0.3 & 0.9 $\pm$ 0.01 \\
\hline
\end{tabular}
\end{minipage}
\begin{minipage}[t]{.45\linewidth}
\vspace{0pt}
\centering
\begin{tabular}{ l  l  c  c }
\multicolumn{4}{c}{T4 DNA}\\
\hline
$c$  & T  & $c/c^{*}$ & $\eta_{0}$ \\  [0.5ex]
\hline
\hline
0.038 & 15 & 0.79 & 5.38 $\pm$ 0.13 \\
\hline
0.023 & 15.7 & 0.58 & 2.43 $\pm$ 0.01 \\
      & 17.3 & 0.72 & 2.33 $\pm$ 0.01 \\
      & 19.4 & 0.85 & 2.23 $\pm$ 0.01 \\
\hline
0.015 & 15.7 & 0.38 & 1.96 $\pm$ 0.01 \\
      & 17.3 & 0.47 & 1.86 $\pm$ 0.01 \\
      & 19.4 & 0.56 & 1.79 $\pm$ 0.01 \\
      & 22   & 0.65 & 1.68 $\pm$ 0.01 \\
      & 24.5 & 0.71 & 1.6 $\pm$ 0.01 \\
\hline
0.0094 & 15 & 0.2  & 1.51 $\pm$ 0.01 \\
       & 20 & 0.36 & 1.48 $\pm$ 0.01 \\
       & 25 & 0.39 & 1.43 $\pm$ 0.01 \\
       & 30 & 0.52 & 1.4 $\pm$ 0.01 \\
       & 35 & 0.59 & 1.37 $\pm$ 0.01 \\
\hline
0.0059 & 15 & 0.12  & 1.36 $\pm$ 0.01 \\
       & 20 & 0.23 & 1.29 $\pm$ 0.01 \\
       & 25 & 0.25 & 1.22 $\pm$ 0.01 \\
       & 30 & 0.33 & 1.16 $\pm$ 0.01 \\
       & 35 & 0.37 & 1.09 $\pm$ 0.01 \\
\hline
0.0038 & 15 & 0.08 & 1.27 $\pm$ 0.01 \\
       & 20 & 0.14 & 1.18 $\pm$ 0.01 \\
       & 25 & 0.15 & 1.09 $\pm$ 0.01 \\
       & 30 & 0.21 & 1.02 $\pm$ 0.01 \\
       & 35 & 0.23 & 0.96 $\pm$ 0.01 \\
\hline
\end{tabular}
\end{minipage}%
\end{table}

\begin{figure}[tbp]
\begin{center}
\begin{tabular}{c}
\resizebox{0.8\linewidth}{!} {\includegraphics*{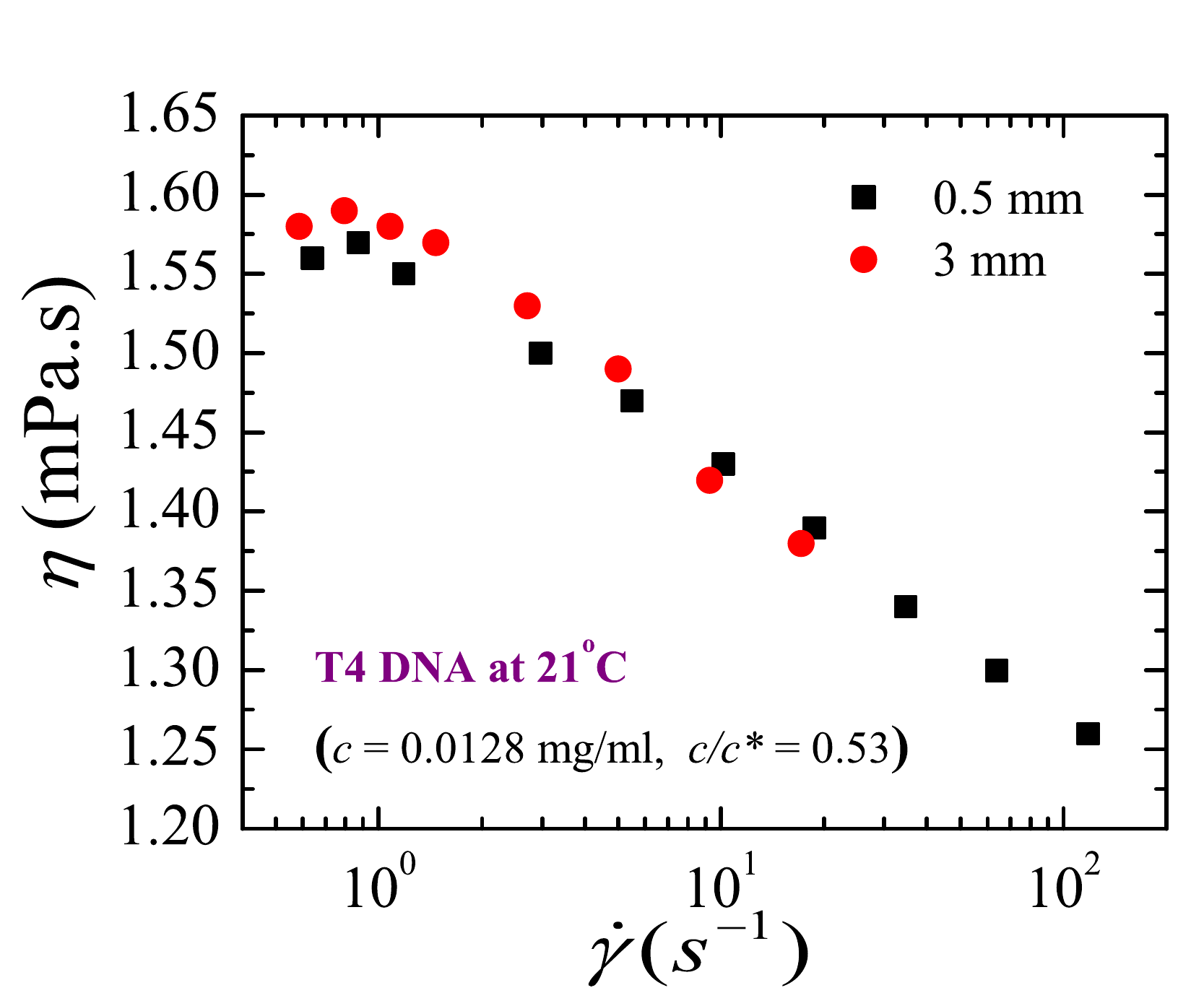}} \\
(a)  \\
\resizebox{0.8\linewidth}{!} {\includegraphics*{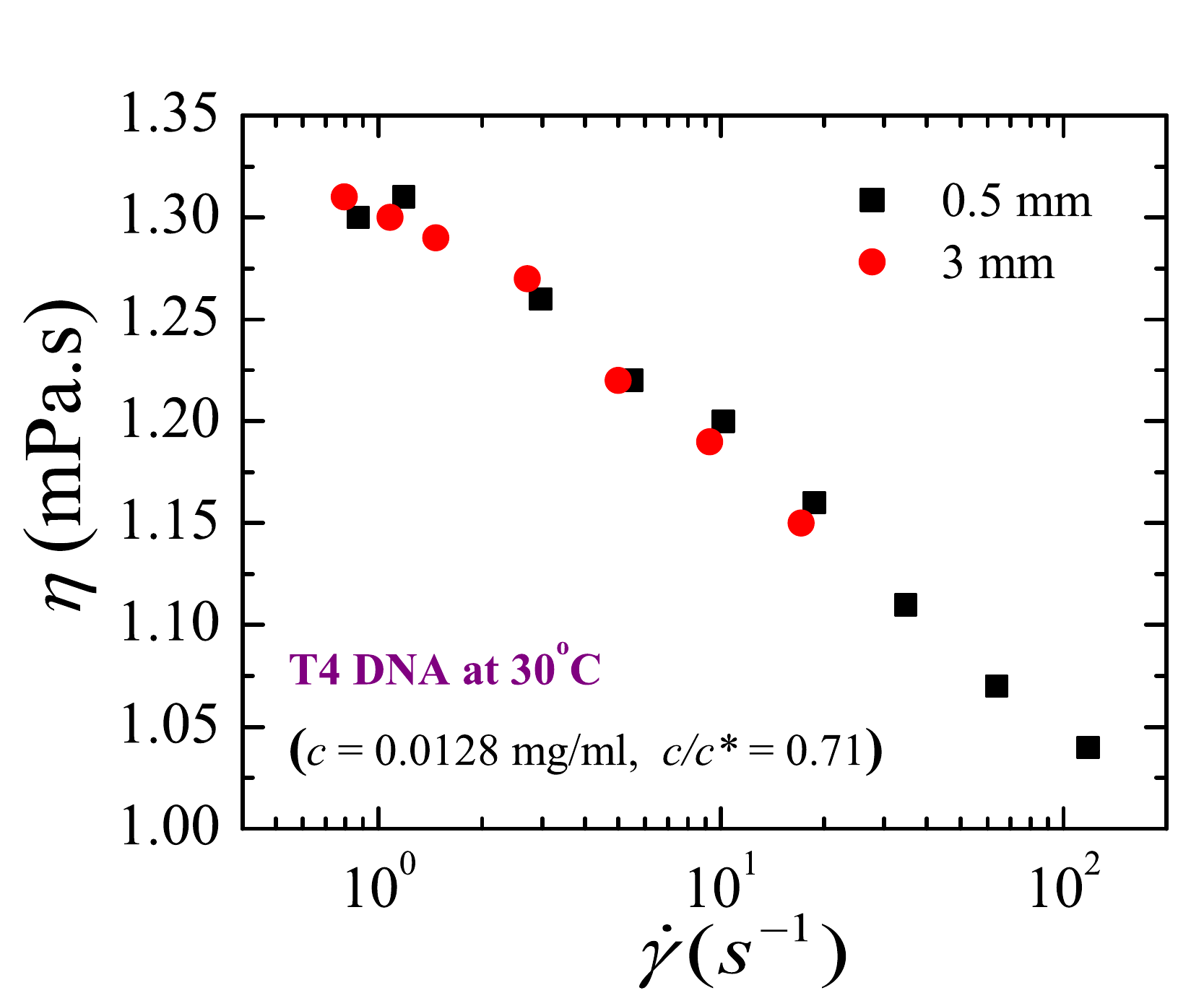}} \\
(b)  \\
\end{tabular}
\end{center}
\caption{\label{fig:gap} Measured solution viscosity $\eta$ as a function of shear rate $\dot \gamma$ for a dilute solution of T4 DNA at two different `gaps' (between the cup and the bob) and at two different temperatures: (a) 21\degC\ and (b) 30\degC. The measurement with the gap of 0.5 mm corresponds to the 1T/1T geometry that has been used for all the measurements in the current work.}
\end{figure}

The shear rate dependence of the measured steady state shear viscosity $\eta$ of the solutions is shown in~\cref{fig:rawvisc}. From the figure, it is clear that the solution viscosity is virtually independent of the shear rate at very low shear rates, which is expected for dilute polymer solutions. The zero shear rate solution viscosities $\eta_{0}$ were determined by least-square fitting of the viscosity values in the plateau region of very low shear rates with a straight line and then extrapolating it to zero shear rate, as shown in the figures. \cref{tab:visc} displays all the zero shear rate viscosities obtained this way for the two molecular weights across the range of concentrations and temperatures examined in the current work. We have also established that the measured viscosity does not depend on rheometer geometry in the range of shear rates employed (in terms of the `gap' between the cup and the bob), by measuring the viscosity of T4 DNA at two different gaps at two different temperatures as shown in~\cref{fig:gap}.

\section{\label{sec:constk} Estimation of the chemistry dependent constant $k$}

The temperature crossover behaviour from $\theta$ solvents to very good solvents for wormlike polymer solutions is described by the solvent quality parameter $\tilde z$, defined by the expression\cite{yamakawa1997}
 \begin{equation}
\tilde z = \left[ \frac{3}{4} K(\lambda L)  \right] z=  \left[ \frac{3}{4} K(\lambda L)  \right]
 k \hat \tau \, \sqrt{M}
 \label{eq:tz}
\end{equation}
where, $\hat \tau=  \left(1- \dfrac{T_{\theta}}{T} \right)$, and $K (\lambda L) = \dfrac{4}{3} - 2.711 \, \dfrac{1}{\sqrt{\lambda L}} + \dfrac{7}{6} \, \dfrac{1}{\lambda L} \, ; \, \text{for}\, \, \lambda L > 6$. The remaining quantities in~\cref{eq:tz} have been defined in the main text. While there is a branch of the function $K  (\lambda L)$ for values of $\lambda L < 6$, we only consider the branch with $\lambda L > 6$, since this is the case for all the DNA considered in this work. Assuming that data can be collapsed onto master plots, the value of $k$ for an experimental system is typically chosen such that experimental and theoretical values of $\tilde z$ agree when the respective equilibrium property values are identical. In the present instance, we compare experimental measurements of the swelling ratios \ah\ and \aeta\ for DNA with the corresponding predictions of Brownian dynamics simulations in order to estimate $k$, as described below.

We assume that the theoretically predicted swelling of any typical property can be represented by the functional form $\alpha = f (\tilde z)$, where, $f (\tilde z) = (1 + a\,\tilde z + b\, \tilde z^{2} + c \, \tilde z^{3})^{m/2}$, with the values of the constants $a$, $b$, $c$, $m$, etc., chosen based on the particular context. This implies that we take the functional dependence of swelling on $\tilde z$ for wormlike chains to be identical to the functional dependence on $z$ for flexible chains. Consider $\alpha^\text{expt}$ to be the experimental value of swelling at a particular value of temperature $T$ and molecular weight $M$. It is then possible to find the Brownian dynamics value of $\tilde z$ that would give rise to the same value of swelling from the expression $\tilde z = f^{-1} (\alpha^\text{expt})$, where $f^{-1}$ is the inverse of the function $f$. Since $\tilde z = \frac{3}{4} K(\lambda L) \, k \, \hat \tau \, \sqrt{M}$, it follows that a plot of $f^{-1}(\alpha^\text{expt})/[\frac{3}{4} K(\lambda L) \, \sqrt{M}]$ versus $\hat \tau$, obtained by  using a number of values of $\alpha^\text{expt}$ at various values of $T$ and $M$, would be a straight line with slope $k$. Once the constant $k$ is determined, both the experimental measurements of swelling and results of Brownian dynamics simulations can be represented on the same plot. Assuming that the $\theta$-temperature is 15\degC\ for the solvent used in this study, we have determined the value of $k$ by following this procedure.

\begin{figure}[!htbp]
\begin{center}
\resizebox{0.9\linewidth}{!} {\includegraphics*{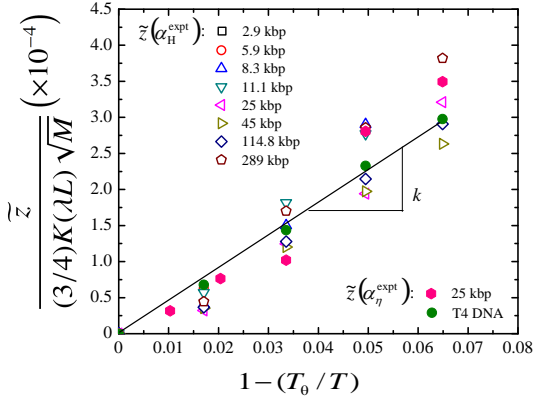}}
\end{center}
\caption{\label{fig:slopek} Determination of the chemistry dependent constant $k$. The data points are least square fitted with a straight line and the slope of this line gives $k$ [Kumar and Prakash, 2003].}
\end{figure}

\cref{fig:slopek} is a plot of $f^{-1}(\alpha^\text{expt})/\sqrt{M}$ versus $\hat \tau$, with measured values of \ah\ and \aeta\ substituted for $\alpha^\text{expt}$, for the various DNA molecular weights considered in this study, and previously by \citet{Pan2014339} Only the temperatures above the theta point are considered here. Values of $\lambda L$ and $K(\lambda L)$ for all the DNA are tabulated in~\cref{tab:z}. The data points were least square fitted with a straight line, and the slope $k$ determined. The value of $k$ found by this procedure is $0.0047 \pm 0.0001$ (g/mol)$^{-1/2}$. Typical values of $\tilde z$, at various $M$ and $T$, obtained by this procedure are reported in~\cref{tab:z}.

\begin{table*}[t]
\addtolength{\tabcolsep}{2pt}
\caption{\label{tab:z} Solvent quality $\tilde{z}$ for DNA at various values of $T$ (in \degC) and $M$. The stiffness parameter ($\lambda^{-1}$) for DNA has been taken to be 1100 \AA, as reported previously in Ref.~\citenum{Yamakawa1984}. Values of $L$ for these DNA have been tabulated previously in \citet{Pan2014339}}
\vspace{2pt}
\centering
\begin{tabular}{c c c c c c c c c}
\hline
  Size & $M$ & $\lambda L$ & $K (\lambda L)$ & \multicolumn{5}{c}{$\tilde{z}$} \\
\cline{5-9}
 (kbp) & ($\times 10^{6}$ g/mol) &  & & 15\degC\ & 20\degC\ & 25\degC\ & 30\degC\ & 35\degC\ \\
\hline
\hline
  2.96  & 1.96 & 9.1   & 0.42 & 0 & 0.05 & 0.09 & 0.14 & 0.18 \\
  5.86  & 3.88 & 18.2  & 0.57 & 0 & 0.09 & 0.18 & 0.26 & 0.34 \\
  8.32  & 5.51 & 27.3  & 0.64 & 0 & 0.12 & 0.24 & 0.35 & 0.46 \\
  11.1  & 7.35 & 36.4  & 0.69 & 0 & 0.15 & 0.29 & 0.43 & 0.57 \\
  25    & 16.6 & 81.8  & 0.79 & 0 & 0.26 & 0.5  & 0.74 & 0.97 \\
  45    & 29.8 & 136.4 & 0.83 & 0 & 0.36 & 0.72 & 1.06 & 1.39 \\
  114.8 & 76   & 354.5 & 0.89 & 0 & 0.62 & 1.23 & 1.81 & 2.38 \\
  165.6 & 110  & 509.1 & 0.91 & 0 & 0.76 & 1.5  & 2.22 & 2.91 \\
  289   & 191  & 890.9 & 0.93 & 0 & 1.03 & 2.03 & 3    & 3.94 \\
\hline
\end{tabular}
\end{table*}

\section{\label{sec:BDSI} Brownian dynamics simulations}

The time evolution of the position vector
${\Vector{r}}_{\mu}(t)$ of bead $\mu$, is described by the non-dimensional stochastic differential equation~\cite{ottinger}
\begin{equation}
\rmd \Vector{r}_{\mu} = \frac{1}{4} \, \sum_{\nu}
\Tensor{D}_{\mu\nu} \cdot \fnu \,
\rmd t + \frac{1}{\sqrt{2}} \, \sum_{\nu} \Tensor{B}_{\mu\nu} \cdot
\rmd \Vector{W}_{\nu}
\label{eq:sde}
\end{equation}
where, the length scale $l_H$ and time scale ${\lambda}_H$ have been used for non-dimensionalization. The dimensionless diffusion tensor
$\bm{\Tensor{D}}_{\mu\nu}$ is a $3 \times 3$ matrix for a fixed pair of beads $\mu$ and
$\nu$. It is related to the hydrodynamic interaction tensor, as
discussed further subsequently. The sum of all the non-hydrodynamic forces on bead $\nu$
due to all the other beads is represented by ${\bm{\Vector{F}}}_{\nu}$, $\bm{\Vector{W}}_{\nu}$ is a Wiener process, and the quantity $\bm{\Tensor{B}}_{\mu\nu }$ is a non-dimensional tensor whose presence leads to multiplicative noise~\cite{ottinger}. Its evaluation requires
the decomposition of the diffusion tensor. Defining the matrices
$\bcal{D}$ and $\bcal{B}$ as block matrices consisting of $N \times N$
blocks each having dimensions of $3 \times 3$, with the $(
\mu,\nu)$-th block of $\bcal{D}$ containing the components of the
diffusion tensor $\bm{\Tensor{D}}_{\mu\nu }$, and the corresponding
block of $\bcal{B}$ being equal to $\bm{\Tensor{B}}_{ \mu\nu}$, the
decomposition rule for obtaining $\bcal{B}$ can be expressed as
\begin{gather}
\bcal{B} \cdot {\bcal{B}}^\textsc{t} = \bcal{D} \label{decomp}
\end{gather}
The non-hydrodynamic forces on a bead $\mu$ are comprised of the non-dimensional spring forces ${\bm{\Vector{F}}}_{\mu}^{\text{spr}}$ and non-dimensional excluded-volume
interaction forces ${\bm{\Vector{F}}}_{\mu}^{\text{exv}}$, \ie,
${\bm{\Vector{F}}}_{\mu} = {\bm{\Vector{F}}}_{\mu}^{\text{spr}} +
{\bm{\Vector{F}}}_{\mu}^{\text{exv}}$. The entropic
spring force on bead $\mu$ due to adjacent beads can be expressed as
${\bm{\Vector{F}}}_{\mu}^{\text{spr}} =
{\bm{\Vector{F}}}^c({\bm{\Vector{Q}}}_{\mu}) -
{\bm{\Vector{F}}}^c({\bm{\Vector{Q}}}_{\mu - 1})$ where
${\bm{\Vector{F}}}^c({\bm{\Vector{Q}}}_{\mu - 1})$ is the force
between the beads $\mu -1$ and $\mu$, acting in the direction of the
connector vector between the two beads ${\bm{\Vector{Q}}}_{\mu - 1} =
{\bm{\Vector{r}}}_{\mu} - {\bm{\Vector{r}}}_{\mu - 1}$. Since simulations are carried out at equilibrium, a linear Hookean spring force is used for modelling the spring forces,
${\bm{\Vector{F}}}^c({\bm{\Vector{Q}}}_{\mu}) = {\bm{\Vector{Q}}}_{\mu}$. The vector $\bm{\Vector{F}}_{\mu}^{\text{exv}}$ is given in
terms of the excluded volume potential $E \left( \bm{\Vector{r}}_{\mu} -
\bm{\Vector{r}}_{\nu} \right)$ between the beads $\mu$ and $\nu$ of the
chain, by the expression,
\begin{equation}
{\bm{\Vector{F}}}_{\mu}^{\text{exv}} = - \sum_{\substack{ \nu = 1\\ \nu \ne \mu}}^N \, \frac{\partial }{ \partial\br_{\mu}} \, E \left(\br_{\mu} - \br_{\nu} \right)
\label{evforce}
\end{equation}
We adopt a narrow Gaussian excluded volume potential in this
work, with $E \left( {\br}_{\mu} - {\br}_{\nu} \right)$ given by,
\begin{equation}
E \left( {\br}_{\mu} - {\br}_{\nu} \right) = \left( \frac{z^* }{
{d^*}^3} \right) \exp \left[ - \frac{ \br_{\mu \nu}^2 }{ {d^*}^2}
\right] \label{evpot}
\end{equation}
where, $\br_{\mu \nu} = \br_{\mu}-\br_{\nu}$, is the vector
between beads $\nu$ and $\mu$, and the parameters $z^*$ and $d^*$
are nondimensional quantities which characterize the narrow
Gaussian potential: $z^*$ measures the strength of the excluded
volume interaction, while $d^*$ is a measure of the range of
excluded volume interaction. The narrow Gaussian potential is a
means of regularizing the Dirac delta potential since it reduces
to a $\delta$-function potential in the limit of $d^*$ tending to
zero.

The non-dimensional diffusion tensor $\bm{\Tensor{D}}_{\nu \mu}$ is related to the non-dimensional hydrodynamic interaction tensor $\bOmega$ through
\begin{equation}
\label{eq:Domega}
{\bm{\Tensor{D}}}_{\mu \nu} = \delta_{\mu \nu} \, \bdelta
+ (1 - \delta_{\mu \nu}) \, \bOmega (\bm{r}_\nu - \bm{r}_\mu)
\end{equation}
where $\bdelta$ and $\delta_{\mu \nu}$ represent a unit
tensor and a Kronecker delta, respectively, while
$\bOmega$ represents the effect of the motion of a
bead $\mu$ on another bead $\nu$ through the disturbances carried by
the surrounding fluid. The hydrodynamic interaction tensor
${\bOmega}$ is assumed to be given by the
Rotne-Prager-Yamakawa (RPY) regularisation of the Oseen function
\begin{equation}
\label{eq:RPY}
{\bOmega}({\bm{\Vector{r}}})
= {\varOmega}_1 \, \bdelta
+ {\varOmega}_2 \frac{{\bm{\Vector{r}}} {\bm{\Vector{r}}}} {{{{r}}}^2}
\end{equation}
where for $r \equiv \vert {\bm{\Vector{r}}} \vert \ge 2 \sqrt{\pi}\hsH$,
\begin{equation}
\label{eq:b1}
{\varOmega}_1 = \frac{3\sqrt{\pi}}{4} \,
\frac{h^{*}}{r} \, \left( 1 + \frac{2\pi}{3} \, \frac{h^{*2}}{r^{2}} \right)
\, \, \, \, \text{and} \, \, \, \,
{\varOmega}_2 = \frac{3\sqrt{\pi}}{4} \, \frac{h^{*}}{r} \, \left( 1 - 2\pi \,
\frac{h^{*2}}{r^{2}} \right)
\end{equation}
while for $0 < r \le 2\sqrt{\pi}\hsH$,
\begin{equation}
\label{eq:b2}
{\varOmega}_1 = 1 - \frac{9}{32} \,
\frac{r}{\hsH\sqrt{\pi}}
\, \, \, \, \text{and} \, \, \, \,
{\varOmega}_2 = \frac{3}{32} \,
\frac{r}{\hsH\sqrt{\pi}}
\end{equation}

In the presence of fluctuating HI, the problem of the computational intensity of calculating the Brownian term is reduced by the use of a Chebyshev polynomial representation for the Brownian term~\cite{fix86,jendrejack:jcp-00}. We have adopted this strategy, and the details of the exact algorithm followed here are given in Ref.~\citenum{PraRavi04}.

\section{Fixman's expressions for $\bar H_{\mu\nu}$
and $\langle \hat{\CS} (t) \rangle_{\mathrm{eq}}$ }
\label{sec:varSI}

Fixman~\cite{Fixman1983} has shown that the equilibrium averaged hydrodynamic interaction tensor is given by
\begin{equation}
  \label{eq:hbar}
   \bar {H}_{\mu\nu}  = \erf(x_{\mu\nu})
  - \frac{1}{\sqrt{\pi}}  \frac{1-\exp({-x_{\mu\nu}^{2}})}{x_{\mu\nu}}
 \end{equation}
\text{where},
\begin{equation}
x_{\mu\nu}  \equiv \sqfrac{2 \, \pi \, h^{*2}}{ \modulus{\mu-\nu}} \quad
  \text{for } \mu\neq\nu
\end{equation}
By defining the components of the $(N-1) \times (N-1)$ matrix $ \widetilde{\bsf{ A}}$, with the expression,
\begin{equation}
  \widetilde{ A}_{jk}  = \sum_{\mu,\nu} \overline{B}_{j\mu} \, H_{\mu\nu} \, \overline{B}_{k\nu}
\end{equation}
where, $ \overline{B}_{k\nu} = \delta_{k+1,\nu} - \delta_{k \nu}, \, \text{for} \, 1 \leq k \leq (N-1) ; \; 1 \leq \nu \leq N$, Fixman~\cite{Fixman1981} has derived the following analytical expression for the stress-stress auto-correlation function of the stochastic process $\brhat_{\nu}$,
\begin{equation}
  \label{eq:csanal}
\langle {\hat \CS (t)} \rangle_{\mathrm{eq}} =   \tr  \left( \exp \left[ -  \frac{1}{2} \, \widetilde{\bsf{ A}} \, \, t \right] \right)
\end{equation}
Clearly, if the RPY tensor is replaced with the Oseen tensor in the definition of $\Tensor{D}_{\mu\nu}$, then $ \widetilde{ A}_{jk}$ is nothing but the modified Kramers matrix~\cite{birdetal2}.

\section{Integration of the correlation functions}

The time correlation function ${\hat \CS}(t)$ is expected to decay as a sum of exponentials~\cite{Fixman1981},
\begin{equation}
{\hat \CS}(t)  = \sum_{k} a_{k} \Exp{-t/\tau_{k}}
\end{equation}
so that,
\begin{equation}
 \Int{0}{\infty}{t} {\hat \CS} (t)  = \sum_{k} a_{k} \, \tau_{k}
\end{equation}
Similar behaviour is expected for $\CS(t)$, although the relaxation spectrum need not be discrete.  We found it sufficient to use a small number of discrete modes (typically three to six
in number) to fit the data with an acceptable error (determined by a
$\chi^{2}$ test of fit). A Levenberg-Marquardt least square
regression algorithm provided as part of GNU-octave package (version
3+) was used to carry out the fitting. Initial guesses for the relaxation
times $\tau_{k}$ have been obtained from estimates of the relaxation spectrum using
the Thurston correlation~\cite{Thurston1974569}.



%
%
%
%

%
%
%

\end{document}